\documentclass[%
 reprint,
 aps,
 prx,
 amsmath,amssymb,
 superscriptaddress,
 floatfix,
]{revtex4-2}
\usepackage{hyperref}
\hypersetup{colorlinks=true,linkcolor=blue,citecolor=blue}
\usepackage[caption=false]{subfig}
\usepackage{graphicx}
\usepackage{dcolumn}
\usepackage{bm}
\usepackage{booktabs}
\usepackage{makecell}

\begin{document}

\preprint{APS/123-QED}

\title{Suppression of Quasiparticle Poisoning to $10^{-11}$ Levels in Superconducting Qubits via Infrared Shielding}

\author{Wei-En Lin}
\affiliation{Research Center for Critical Issues, Academia Sinica, Tainan, 711010, Taiwan}
\affiliation{Department of Physics, National Central University, Taoyuan, 320317, Taiwan}

\author{Chen-Hsun Ma}
\affiliation{Research Center for Critical Issues, Academia Sinica, Tainan, 711010, Taiwan}
\affiliation{Department of Physics, National Taiwan University, Taipei, 106319, Taiwan}

\author{Li-Chieh Hsiao}
\affiliation{Research Center for Critical Issues, Academia Sinica, Tainan, 711010, Taiwan}
\affiliation{Department of Physics, National Cheng Kung University, Tainan, 701401, Taiwan}

\author{Erh-Hsiang Yeh}
\affiliation{Department of Physics, National Central University, Taoyuan, 320317, Taiwan}

\author{Wei-Lun Peng}
\affiliation{Department of Physics, National Changhua University of Education, Changhua, 500207, Taiwan}
 
\author{Yu-Sen Wei}
\affiliation{Research Center for Critical Issues, Academia Sinica, Tainan, 711010, Taiwan}
\affiliation{Department of Physics, National Central University, Taoyuan, 320317, Taiwan}

\author{Hsi-Sheng Goan}
\affiliation{Department of Physics, National Taiwan University, Taipei, 106319, Taiwan}
\affiliation{Center for Quantum Science and Engineering, National Taiwan University, Taipei, 106319, Taiwan}
\affiliation{Physics Division, National Center for Theoretical Sciences, Taipei, 106319, Taiwan}

\author{Cen-Shawn Wu}
\affiliation{Research Center for Critical Issues, Academia Sinica, Tainan, 711010, Taiwan}
\affiliation{Department of Physics, National Changhua University of Education, Changhua, 500207, Taiwan}

\author{Chung-Ting Ke}\email{ctke@gate.sinica.edu.tw}
\affiliation{Research Center for Critical Issues, Academia Sinica, Tainan, 711010, Taiwan}
\affiliation{Institute of Physics, Academia Sinica, Taipei, 115201, Taiwan}

\author{Yung-Fu Chen}\email{yfuchen@cc.ncu.edu.tw}
\affiliation{Research Center for Critical Issues, Academia Sinica, Tainan, 711010, Taiwan}
\affiliation{Department of Physics, National Central University, Taoyuan, 320317, Taiwan}
\affiliation{Center of High Energy and High Field Physics, National Central University, Taoyuan, 320317, Taiwan}
\affiliation{Quantum Technology Center, National Central University, Taoyuan, 320317, Taiwan}
\affiliation{Taiwan Semiconductor Research Institute, National Institutes of Applied Research, Hsinchu, 300091, Taiwan}
\affiliation{National Center for Excellence in Quantum Information Science and Engineering, National Tsing Hua University, Hsinchu, 300044, Taiwan }

\author{Chii-Dong Chen}
\affiliation{Research Center for Critical Issues, Academia Sinica, Tainan, 711010, Taiwan}
\affiliation{Institute of Physics, Academia Sinica, Taipei, 115201, Taiwan}

\begin{abstract}
Quasiparticle poisoning is a bottleneck for superconducting qubits, limiting their coherence and the scalability of quantum processors. Stray infrared (IR) radiation is one of its most pervasive sources, yet the effectiveness of individual shielding elements and the scaling with the number of coaxial lines have rarely been systematically quantified. We address both for floating transmons with large capacitor pads by removing the shielding elements one at a time. In the best configuration, the parity-switching rate falls to about $0.1$\,Hz, corresponding to an upper estimate of the resident quasiparticle density per Cooper pair of $4.11\times10^{-11}$, the lowest value reported to date. At this density, the quasiparticle-induced relaxation rate is more than three orders of magnitude below the measured relaxation rate. Using the simulated antenna coupling of the qubits, we convert the measured rates into an effective blackbody temperature of the radiation at the device, which allows shielding performance to be compared across experiments. This temperature is bounded to below $230$\,mK, equivalent to an attenuation of 4\,K radiation of at least $96$\,dB. We find that the mixing-chamber shield, coated with an IR absorber, provides the dominant suppression. In-line IR filters reduce the contribution of each coaxial line from $0.1$--$0.2$\,Hz to below $10^{-7}$\,Hz, so that the number of lines no longer limits the rate as quantum processors scale up. Sealing the enclosure with indium further lowers the rate and the effective temperature of the qubits. With all elements in place, the rate continues to decrease over weeks after cooldown, indicating that the remaining parity switching originates from a time-dependent source rather than from stationary radiation. These findings provide practical design guidelines for scaling up superconducting processor packages and implementing IR shielding in superconducting detectors.
\end{abstract}

\maketitle

\section{\label{sec:level1}Introduction}

A quasiparticle (QP) is an elementary excitation in a superconductor, arising from the breaking of a Cooper pair \cite{tinkham2004introduction}. Pair-breaking events directly impact the performance of superconducting devices. In single-photon detectors and kinetic inductance detectors, QP excitation increases the dark count, thereby reducing the detector's sensitivity \cite{pankratov2025detection,walsh2021josephson,PhysRevApplied.7.014012,alesini2020development,PhysRevX.14.041005}. Similarly, in superconducting qubits, nonequilibrium QPs are known to be a major source of decoherence, limiting the energy relaxation time $T_1$ and contributing to residual excited-state population $P_1$ \cite{shaw2008kinetics, martinis2009energy, catelani2011quasiparticle, catelani2011relaxation, yan2016flux, wilen2021correlated}. Furthermore, their generation by high-energy particles induces spatially correlated errors that severely undermine quantum error correction across multi-qubit arrays~\cite{martinis2021saving, mcewen2022resolving, xu2022distributed, google2023suppressing}. Suppressing these dynamics is therefore a fundamental prerequisite for realizing scalable superconducting quantum processors.    Extensive studies have shown that the QP density per Cooper pair, $x_\mathrm{qp}=n_\mathrm{qp}/n_\mathrm{cp}$, typically ranges from $10^{-6}$ to $10^{-9}$ in superconducting devices, exceeding the expected thermal equilibrium value by tens of orders of magnitude at millikelvin temperatures \cite{shaw2008kinetics, martinis2009energy, de2011number, wang2014measurement, serniak2018hot, liu2024quasiparticle}. Interestingly, while their density is highly nonequilibrium, recent measurements indicate that the QP energy distribution itself can remain close to thermal equilibrium \cite{connolly2024coexistence}. Understanding the origin of this excess QP population is therefore a central challenge for improving qubit coherence.

A variety of mechanisms have been identified as sources of nonequilibrium QPs, along with corresponding mitigation strategies. For example, measurement-induced effects, due to excessive drive power or unintended microwave leakage, can generate QPs \cite{wang2014measurement}, but these contributions can typically be minimized through careful calibration of the measurement setup. On the other hand, high-energy particle impacts, such as cosmic rays, represent another important source of QPs \cite{wilen2021correlated, mcewen2022resolving}. These events generate energetic phonons that propagate through the substrate and subsequently break Cooper pairs and create correlated errors in qubit arrays \cite{mcewen2022resolving, martinis2021saving,wilen2021correlated,Vepsalainen2020,harrington2025synchronous,li2025cosmic}. Mitigation strategies include the use of heavy shielding materials and underground operation \cite{cardani2021reducing, Bratrud:2024qnk}, and phonon engineering techniques such as phonon trapping \cite{henriques2019phonon} and phonon down-conversion \cite{martinis2021saving,Iaia2022PhononDT}. More recently, microscopic phonon bursts associated with stressed materials in device fabrication or packaging have been observed \cite{anthony2024stress, yelton2025correlated}, although their microscopic origin remains unclear. In addition, a gradual reduction of the QP tunneling rate over weeks after cooldown has been attributed to thermal radiation from weakly anchored components that continue to cool inside the cryostat \cite{2026assessing}. Meanwhile, approaches such as gap engineering to spatially separate QPs from Josephson junctions \cite{sun2012measurements, nho2026recovery} and the use of vortices as QP traps \cite{wang2014measurement,PhysRevB84220502} further illustrate the range of techniques being explored to control QP dynamics.

A more pervasive source of QPs is stray infrared (IR) radiation, which can be absorbed by the chip or qubit and then break Cooper pairs \cite{liu2024quasiparticle, pan2022engineering}. The role of stray IR radiation was established in early single-electron devices. Photon-activated switching was identified in a superconducting single-electron transistor~\cite{PhysRevB519407}, and an indium-sealed, double-shielded sample stage was shown to dramatically reduce QP excitation events~\cite{PhysRevB85012504}. Building on these foundations, significant progress has been made in suppressing this mechanism through improved light-tight shielding \cite{barends2011minimizing, kreikebaum2016optimization, connolly2024coexistence}, infrared filtering \cite{danilin2022engineering, nho2026recovery}, and absorptive coatings inside radiation shields and sample enclosures \cite{barends2011minimizing, kreikebaum2016optimization}. Device-level strategies, such as reducing qubit dimensions, have also been shown to decrease resonant absorption of IR photons \cite{rafferty2021spurious, liu2024quasiparticle, pan2022engineering}.

Among these mechanisms responsible for QP generation, stray IR radiation has emerged as one of the dominant and most controllable sources. Consequently, extensive radiation shielding is commonly employed to suppress nonequilibrium QPs and improve qubit coherence. Ref.~\cite{malevannaya2025engineering} provides comprehensive guidelines for designing shielding systems that protect quantum processors from both IR radiation and stray electromagnetic fields. However, implementing multiple layers of IR shielding introduces practical challenges. Specifically, each shielding stage must be properly thermalized. Inadequate thermal anchoring can lead to elevated shield temperatures, resulting in excess blackbody radiation that undermines the intended suppression of QPs. Furthermore, while an indium seal is used to ensure light-tightness, indium is a superconductor with low thermal conductivity at millikelvin temperatures. A strategic arrangement of the shield is therefore crucial to avoid weak thermal anchoring while maintaining a light-tight environment.

Beyond these practical challenges, several questions relevant to scaling remain open. Few studies have quantified the contribution of each element of a shielding assembly, or the dependence of the residual radiation on the number of coaxial lines, which grows with processor size. It also remains unclear what limits the QP density once stray radiation is suppressed. In this work, we address these questions with floating transmons whose large capacitor pads make them sensitive antennas for pair-breaking radiation~\cite{rafferty2021spurious, liu2024quasiparticle, pan2022engineering}. We measure the QP tunneling rate across the junction, commonly referred to as the parity-switching rate~\cite{sun2012measurements, riste2013millisecond}, in a multi-layer shielding assembly thermalized to the mixing-chamber (MXC) plate, and compare configurations in which the MXC shield, the in-line IR filters, or the indium seal is removed individually.

In the best-shielded configuration, the parity-switching rate falls to about $0.1$\,Hz and continues to decrease over weeks after cooldown. Stray photons can also switch the parity directly through photon-assisted parity switching (PAPS), in which a pair-breaking photon is absorbed at the junction~\cite{houzet2019,diamond2022distinguishing}. Assuming this contribution to be small, as suggested by the continued decrease of the rate, we obtain an upper estimate of the resident QP density, $x_{\mathrm{qp}}\approx4.11\times10^{-11}$, the lowest value reported to date, at which the QP-induced relaxation rate is more than three orders of magnitude below the measured relaxation rate. Expressing the measured rates in terms of radiation, we bound the effective blackbody temperature at the device to below $230$\,mK, equivalent to an attenuation of 4\,K radiation of at least $96$\,dB. Removing individual elements shows that the MXC shield provides the dominant suppression and that the in-line IR filters remove the contribution of the coaxial lines, so that the number of lines no longer limits the parity-switching rate once every line is filtered. Without the indium seal of the enclosure, the parity-switching rate saturates at $2$--$4$ times the lowest rate reached with the seal, and the effective qubit temperature rises by $6$--$9$\,mK relative to the best-shielded configuration. With all elements in place, the continued decrease of the rate indicates that the remaining parity switching arises from a time-dependent source rather than from sources that are constant in time, such as a stationary photon background, environmental radioactivity, or cosmic rays~\cite{mannila2022}. Possible origins of this source are stress relaxation~\cite{anthony2024stress,yelton2025correlated} and slowly cooling components~\cite{2026assessing}.

\section{\label{sec:level2}Experimental setup}\label{sec2}

\begin{figure}
\centering
\includegraphics[width=\columnwidth]{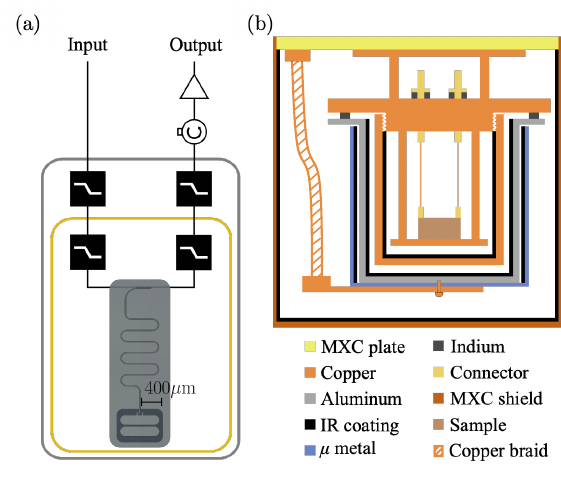}
\caption{Experimental setup schematics of Config.~A. (a) Qubit measurement wiring. The device image shows Q2, one of the floating transmon qubits, coupled to a readout resonator, which is in turn coupled to an open transmission line. The chip is enclosed within a dedicated multi-layer shielding assembly (yellow box) and the MXC radiation shield provided by Bluefors (gray box). Two IR filters are installed on each of the input and output lines. (b) Detailed schematic of shielding Config.~A. The shielding structure is concentrically arranged from inside to outside as follows: a copper enclosure, an aluminum enclosure, and a mu-metal shield. The copper enclosure is sealed with a threaded copper top plate. The aluminum enclosure is sealed to the same top plate using an indium ring compressed by screws. The SMA feedthroughs mounted on the copper top plate are also sealed with indium rings. The interior surfaces of all three shields, as well as the MXC shield, are coated with an IR-absorbing paste.}
\label{fig:FIG1}
\end{figure}

We begin by discussing the most comprehensive multi-layer shielding setup, Config.~A. Figure \ref{fig:FIG1}(a) illustrates the qubit measurement wiring along with a device image. The device chip contains four floating, single-junction transmon qubits, three of which are operational and labeled Q1, Q2, and Q3. Q1 and Q3 share one pad geometry, and Q2 has another (Table~\ref{table:TABLE1}).  The qubit transition frequencies $\omega_{01}/2\pi$ are approximately 3$\,$GHz. The large-scale structures, including the resonator and wiring, are fabricated from 100$\,$nm sputtered aluminum and patterned by dry etching. The qubit pads and Josephson junctions are defined using electron-beam lithography, followed by aluminum deposition and a lift-off process in a single fabrication step. Each qubit incorporates a Manhattan-style Josephson junction with nominal dimensions of 150$\,$nm $\times$ 100$\,$nm, consisting of a 50$\,$nm-thick bottom electrode and a 100\,nm-thick top electrode. The two capacitor pads have dimensions of 750$\,$$\mu$m $\times$ 175 (150)$\,$$\mu$m and are separated by a 30 (80)$\,$$\mu$m gap. The pads are positioned approximately 60--110 (30--110)$\,$$\mu$m away from the surrounding ground plane for Q2 (Q1 and Q3), respectively. This large-gap qubit design reduces the energy participation ratio of lossy interfaces. However, at these dimensions, the pad structure acts as an effective dipole antenna for absorbing IR radiation \cite{rafferty2021spurious} at frequencies $\sim$100$\,$GHz and above, beyond the pair-breaking threshold of aluminum (see Sec.~II of the Supplemental Material~\cite{SM}\nocite{openEMS,hmmlearn,ambegaokar1963} for details). As a result, absorbed photons can break Cooper pairs, making the qubit sensitive to incident IR radiation. 

\begin{table*}[t]
\centering
\caption{Qubit device parameters.}
\label{table:TABLE1}
\renewcommand{\arraystretch}{1.2} 
\setlength{\tabcolsep}{15pt}
\begin{ruledtabular}
\begin{tabular}{cccc}
Parameter & Q1 & Q2 & Q3\\
\hline
Pad length   ($\mu$m) &  $750$ & $750$ & $750$ \\
Pad width  ($\mu$m) &  $150$ & $175$ & $150$ \\
Inter-pad gap ($\mu$m) &  $80$ & $30$ & $80$ \\
Pad-to-ground gap  ($\mu$m) &  $30$--$110$ & $60$--$110$ & $30$--$110$ \\
Bare resonator frequency $\omega_{r}/2\pi$ (GHz) & 4.912 & 5.012 & 5.113\\
Qubit transition frequency $\omega_{01}/2\pi$ (GHz) & 3.077 & 2.946 & 3.359 \\
Qubit-resonator coupling strength $g/2\pi$ (MHz) & 25.7 & 15.4 & 16.5 \\
Resonator photon decay rate  $\kappa/2\pi$ (MHz) & 0.507 & 0.403 & 0.333 \\
Anharmonicity $\alpha/2\pi$ (MHz) & -239 & -210 & -235 \\
Josephson energy $E_J/h$ (GHz) & 6.80 & 6.91 & 7.98 \\
Charging energy $E_C/h$ (GHz) & 0.199 & 0.178 & 0.199 \\
$E_J/E_C$ & 34.2 & 38.8 & 40.1 \\
Purcell-limited lifetime $T^{\rm{pur}}_1=\frac{1}{\kappa}(\frac{\omega_{01}-\omega_{r}}{g})^2$ (ms) &  1.6 & 7.1 & 5.4 \\
\end{tabular}
\end{ruledtabular}
\end{table*}

Each qubit is capacitively coupled to an individual $\lambda/4$ resonator at around 5$\,$GHz for dispersive readout. The resonators are connected to a common readout feedline, enabling multiplexed readout. Both microwave control signals and readout signals are delivered through this shared feedline. The Purcell-limited lifetime $T_1^{\rm{pur}}$ set by the measurement line is on the order of 1--10$\,$ms. The qubits operate in the transmon regime with a Josephson-to-charging energy ratio $E_J/E_C\approx34$--$40$, placing them in a moderately charge-sensitive regime compared to conventional large $E_J/E_C$ transmons \cite{koch2007charge}. This design choice enhances their detectability to parity-switching events \cite{riste2013millisecond}. The parameters of the qubits studied in this work are summarized in Table \ref{table:TABLE1}.

The chip is wire-bonded to a printed circuit board (PCB), which is enclosed in a copper housing. Electrical connections are made via pogo-pin contacts to the PCB and routed to coaxial cables. To suppress environmental IR radiation and magnetic noise, the chip is enclosed within a dedicated multi-layer shielding assembly (indicated by the yellow box in Fig. \ref{fig:FIG1}(a)) and the MXC radiation shield provided by Bluefors (gray box). Figure \ref{fig:FIG1}(b) presents a detailed schematic of the shielding configuration implemented around the MXC plate in this work. The shielding structure is arranged concentrically from inside to outside as follows: a copper enclosure, an aluminum enclosure, and a mu-metal shield, with typical diameters and heights of approximately 12$\,$cm and 16$\,$cm, respectively. The two innermost enclosures (copper and aluminum) are designed to be light-tight to minimize IR photon penetration. The copper enclosure is sealed with a threaded copper top plate to ensure robust mechanical contact and electrical continuity while minimizing IR leakage. The aluminum enclosure is sealed against the same top plate using an indium ring compressed by screws, forming a light-tight interface \cite{connolly2024coexistence,PhysRevB85012504}. To further suppress IR leakage through the measurement wiring, the SMA feedthroughs mounted on the top plate are also sealed with indium rings to eliminate gaps that could act as IR leakage channels \cite{connolly2024coexistence}. Additionally, the interior surfaces of all three shields, as well as the MXC shield, are coated with an IR-absorbing paste \cite{mcdermott2022private} to absorb and reduce stray IR radiation.

All shielding components, together with the sample holder, are carefully thermally anchored to the MXC plate to ensure proper thermalization at base temperature. Specifically, as illustrated in Fig. \ref{fig:FIG1}(b), the mu-metal shield is secured via a brass screw to both the aluminum enclosure and a copper strip. This arrangement provides a high-thermal-conductivity path to the MXC plate through a flexible copper braid. Maintaining all shielding stages near the base temperature reduces residual IR emission and suppresses IR-induced QP generation in the qubits.

In addition, IR filters (Sweden Quantum, HERD-2) are installed on the input and output lines to prevent pair-breaking photons from propagating to the chip through the coaxial lines. Two filters are placed on each line, one inside the shielding assembly and one outside, as shown in Fig.~\ref{fig:FIG1}(a). This dual-stage filtering further reduces nonequilibrium QP generation induced by stray radiation. The full wiring, including the attenuation and filtering at each stage, is described in Sec.~I of the Supplemental Material~\cite{SM}.

\section{Parity-switching measurement}
\label{sec:parity}

\begin{figure}
\centering
\includegraphics[width=\columnwidth]{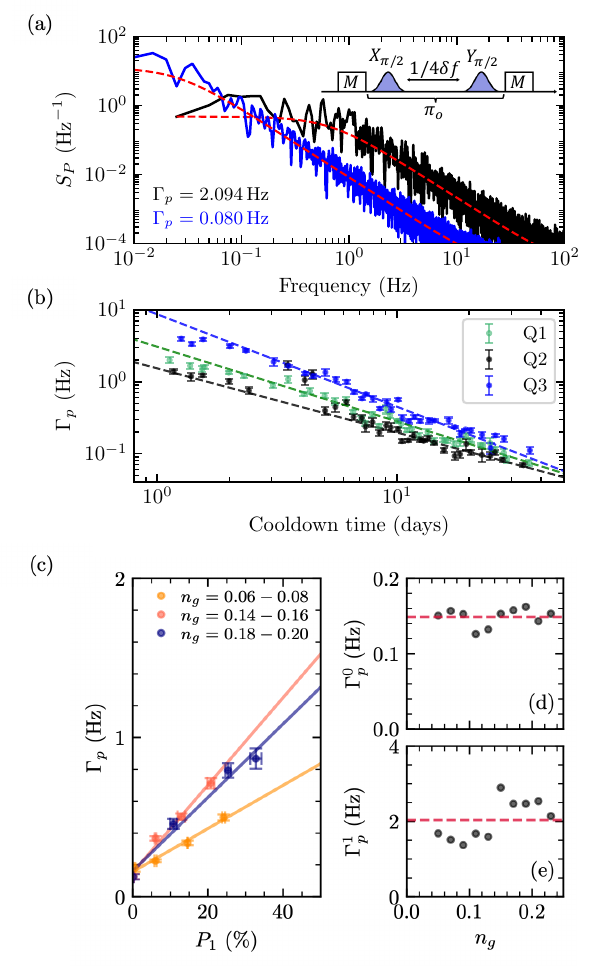}
\caption{Parity-switching measurements. (a) PSD of the parity-switching signal for Q2 measured on Days~1 (black) and 33 (blue), each averaged over five traces. Each spectrum is fitted with a Lorentzian (dashed lines), yielding $\Gamma_p=2.09$\,Hz (Day~1) and $0.080$\,Hz (Day~33). Inset: pulse sequence mapping the charge parity onto the qubit state, where $M$ denotes readout and $\pi_{o}$ indicates that the sequence acts as a $\pi$ pulse only for odd parity. (b) Extracted $\Gamma_p$ vs. time for all three qubits over 36 days, fitted to a power law with exponents $-1.04$, $-0.89$, and $-1.28$ for Q1--Q3, respectively. Error bars denote the standard deviation of the mean over repeated measurements. (c) $\Gamma_p$ vs. $P_1$ for Q2 at several gate charges $n_g$ (Day~15), from which $\Gamma_p^0$ and $\Gamma_p^1$ are extrapolated. (d) $\Gamma_p^0$ and (e) $\Gamma_p^1$ as a function of $n_g$, extracted as in (c). Red dashed lines show the $n_g$ dependence predicted by Fermi's golden rule, with $x_{\mathrm{qp}}$ and $\delta\Delta$ determined from the $n_g$-averaged rates (Sec.~IV of the Supplemental Material~\cite{SM}).}
\label{fig:FIG2}
\end{figure}

We measure QP-induced parity-switching events in all three qubits under shielding Config.~A.  All data in Figs.~\ref{fig:FIG2}--\ref{fig:FIG4} were obtained in Config.~A. The charge-parity states are mapped onto the qubit computational states $|0\rangle$ and $|1\rangle$ using a Ramsey-like pulse sequence with a free-evolution time $\tau=1/4\delta f$, where $2\delta f$ is the frequency difference between the two parity states~\cite{riste2013millisecond}. The measurement sequence is illustrated in the inset of Fig. \ref{fig:FIG2}(a). The parity state is sampled by repeating the sequence every 4--20$\,$ms over a total acquisition time of 100--400$\,$s, generating a time series of parity outcomes. The average single-shot parity measurement fidelity is approximately 0.85, primarily limited by the excited-state readout fidelity of the qubit. To mitigate errors arising from imperfect mapping fidelity, the recorded parity time traces are analyzed using a Hidden Markov Model, which reconstructs the most probable underlying parity sequence \cite{rabiner1989hmm} (see Sec.~III of the Supplemental Material~\cite{SM} for details). The corrected time series is then Fourier transformed to obtain the power spectral density (PSD). The spectrum is fitted to a Lorentzian function, from which the parity-switching rate $\Gamma_p$ is extracted. 

Figure \ref{fig:FIG2} presents the results of the parity-switching measurements. Figure \ref{fig:FIG2}(a) shows the PSD of the parity-switching signal for Q2 measured on Days 1 (black) and 33 (blue), where time is referenced to when the MXC plate reaches 20$\,$mK. Each PSD is obtained by averaging 5 raw spectra to improve the signal-to-noise ratio. The spectra are well fitted by a Lorentzian function. The fitted $\Gamma_p$ are 2.09$\,$Hz on Day~1 and 0.080$\,$Hz on Day~33, indicating a significant reduction over time.

Figure \ref{fig:FIG2}(b) summarizes the extracted $\Gamma_p$ as a function of time for all three qubits over a 36-day measurement period. For each qubit, $\Gamma_p$ decreases monotonically with time. The decay roughly follows a power-law dependence, with an exponent close to $-1$ for Q1 and Q2, and $-1.3$ for Q3. By Day~36, $\Gamma_p$ reaches $0.07$--$0.11$\,Hz (Table~\ref{table:TABLE2}).

The observed long-term decrease in $\Gamma_p$ implies that QP generation is no longer dominated by a stationary IR background incident on the chip at this level of shielding~\cite{mannila2022}. Together with the low value of $\Gamma_p$, this indicates that the IR shielding is highly effective, as quantified in Sec.~\ref{sec:config}. This result is notable because the large-pad qubits used in this work are expected to absorb IR radiation more efficiently than transmon qubits with small capacitor gaps and pads~\cite{liu2024quasiparticle,rafferty2021spurious}. The remaining parity switching more likely arises from a time-dependent source, such as phonon bursts associated with stress relaxation in nearby thin-film materials~\cite{anthony2024stress}, for which a power-law decay with an exponent of approximately $-0.7$ has been reported~\cite{yelton2025correlated}. A similar decrease has alternatively been attributed to thermal radiation from weakly anchored components that cool slowly inside the cryostat~\cite{2026assessing}, and our data do not distinguish between these two possibilities. We note, however, that the power-law exponent extracted in this work is stronger in magnitude than the value of $-0.7$ quoted above, although comparable to the value of $-0.9$ observed in a superconducting island~\cite{mannila2022}. The faster decay observed in our devices may originate from differences in materials, fabrication processes, or the stress-relaxation and thermalization dynamics of the surrounding sample holder structures.

The excited-state parity-switching rate $\Gamma_p^1$ can be used to estimate the QP density $x_{\mathrm{qp}}$ through the relation \cite{Glazman2021Bogoliubov}
\begin{equation}
    \Gamma_p^1 = \frac{16E_J}{\hbar\pi}\sqrt{\frac{E_C}{8E_J}}\sqrt{\frac{\Delta}{2\hbar\omega_{01}}}x_{\mathrm{qp}}, 
    \label{eq:Gamma1}
\end{equation}
where $\Delta$ is the superconducting gap and $\hbar$ is the reduced Planck constant. $\Gamma_p^1$ is related to $\Gamma_p$ through $\Gamma_p = \Gamma_p^0(1 - P_1) + \Gamma_p^1 P_1$, where $\Gamma_p^0$ denotes the ground-state parity-switching rate. Both rates are extracted by measuring $\Gamma_p$ as a function of the excited-state population $P_1$ and extrapolating to $P_1=0$ and $1$, respectively. $P_1$ is controlled using variable-amplitude qubit-scrambling pulses applied every 10$\,$$\mu$s between successive parity-state measurements \cite{diamond2022distinguishing,connolly2024coexistence,nho2026recovery}. Figure~\ref{fig:FIG2}(c) shows $\Gamma_{p}$ as a function of $P_1$ for Q2 at several gate charges $n_{g}$, measured on Day~15. The values of $\Gamma_p^0$ and $\Gamma_p^1$ extracted in this way are shown versus $n_{g}$ in Figs.~\ref{fig:FIG2}(d) and \ref{fig:FIG2}(e). Their averages over $n_g$ are $\bar\Gamma_p^0=0.149$\,Hz and $\bar\Gamma_p^1=2.04$\,Hz. Using Eq.~(\ref{eq:Gamma1}) with $\Gamma_p^1=\bar\Gamma_p^1$, $\Delta = 186.5\,\mu\text{eV}$, obtained from the measured critical temperature $T_c = 1.23$\,K, and $\omega_{01}/2\pi \approx 2.946$\,GHz, we estimate a QP density of $x_{\mathrm{qp}} \approx 5.86 \times 10^{-11}$. Equation~(\ref{eq:Gamma1}) assumes equal superconducting gaps on the two sides of the junction. Since the gap of aluminum increases as the film is made thinner~\cite{Court_2008}, the two electrodes of our junction differ in gap. From the ratio $\bar\Gamma_p^1/\bar\Gamma_p^0$, we extract a gap difference of $\delta\Delta/h\approx0.71$\,GHz (Sec.~IV.A of the Supplemental Material~\cite{SM}). Including it raises the estimate of the QP density in the low-gap film to $x_{\mathrm{qp}}\approx8.93\times10^{-11}$, about seven times that in the high-gap film. Unlike in gap-engineered devices, in which $\delta\Delta>\hbar\omega_{01}$ suppresses QP tunneling in $|1\rangle$~\cite{nho2026recovery}, here $\delta\Delta<\hbar\omega_{01}$, so that the low $\Gamma_p^1$ results from a low QP density rather than from a suppressed tunneling channel. Furthermore, as shown in Fig.~\ref{fig:FIG2}(b), the measured parity-switching rate $\Gamma_{p}$ gradually decreases over time. Assuming $\bar\Gamma_p^1 \propto \Gamma_{p}$, this trend indicates a corresponding reduction in $x_{\rm{qp}}$. By the final measurement day (Day~36), $\Gamma_{p}$, and therefore $\bar\Gamma_p^1$, decreases by $\sim 54\%$ relative to Day~15, implying that $x_{\rm{qp}}$ reaches $4.11 \times 10^{-11}$. To the best of our knowledge, this is the lowest QP density reported in the literature to date.

Both $\Gamma_p^1$ and $\Gamma_p^0$ vary with the gate charge $n_g$, as shown for Q2 in Figs.~\ref{fig:FIG2}(d) and \ref{fig:FIG2}(e) and for all three qubits in Sec.~IV.B of the Supplemental Material~\cite{SM}. This behavior contrasts with previous reports, which found $\Gamma_p(P_1)$ to be independent of $n_g$~\cite{connolly2024coexistence}. We evaluate the QP tunneling rate with charge-parity matrix elements computed exactly by numerical diagonalization, rather than in the transmon limit~\cite{catelani2014parity,marchegiani2022asymmetric}. The predicted rates, depicted by the red dashed lines in Figs.~\ref{fig:FIG2}(d) and \ref{fig:FIG2}(e), have relative variations below $2\times10^{-5}$ over the measured range, so that the observed dependence does not originate from the tunneling matrix elements. The dependence may instead arise from leakage, induced by the scrambling pulses or by measurement-induced state transitions (MIST) during readout, into higher transmon levels whose charge dispersion is orders of magnitude larger than that of $|1\rangle$ and out of which QPs tunnel faster~\cite{Dumas2024prx,fechant2025offset,WeiEnMIST}. Since MIST sets in at values of $n_g$ that depend on the initial state~\cite{fechant2025offset}, this contribution can produce abrupt changes at different $n_g$ in $\Gamma_p^0$ and $\Gamma_p^1$. Relaxation and MIST during readout also bias the extraction of these rates through $P_1$, the latter by an amount that varies with $n_g$ (Sec.~IV.B of the Supplemental Material~\cite{SM}). The uncertainty of $\delta\Delta$, in contrast, hardly affects the QP density. Since $\delta\Delta<\hbar\omega_{01}$, QP tunneling in $|1\rangle$ is not thermally activated. For example, $x_{\mathrm{qp}}$ for Q2 changes by less than $10\%$ for $\delta\Delta/h$ between $0.5$ and $1.2$\,GHz and remains at the $10^{-11}$ level even for $\delta\Delta=0$ (Sec.~IV.A of the Supplemental Material~\cite{SM}).

\section{Effective qubit temperature and energy relaxation time}
\label{sec:Tq}

\begin{figure}
\centering
\includegraphics[width=\columnwidth]{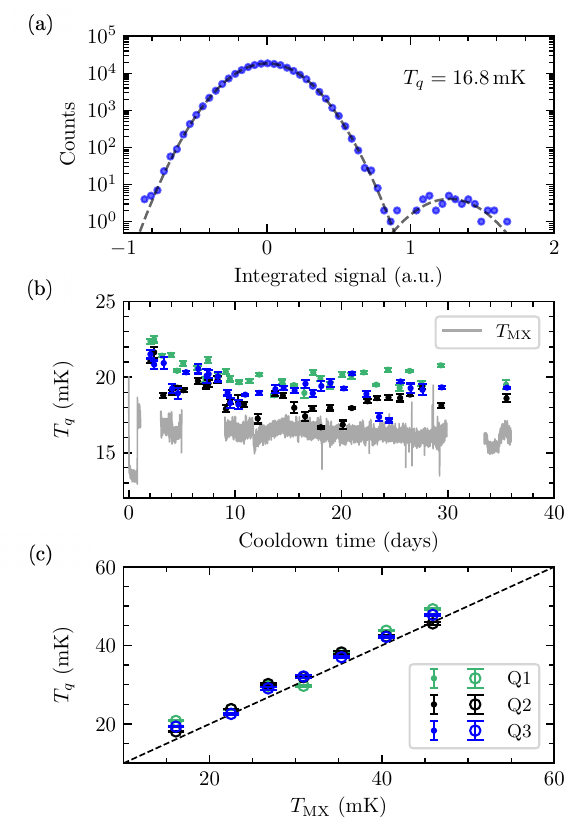}
\caption{Effective qubit temperature. (a) Qubit state readout for Q2 on Day~20, yielding $P_1 = 0.0218\%$ and a corresponding $T_{q}$ = 16.8$\,$mK. (b) $T_{q}$ vs. time for all three qubits, with $T_{\rm{MX}}$ shown for comparison. (c) $T_{q}$ as a function of $T_{\rm{MX}}$ for all three qubits. The legend in (c) applies to both (b) and (c). The dashed line indicates $T_{q} = T_{\rm{MX}}$.}
\label{fig:FIG3}
\end{figure}

Next, we examine the effective qubit temperature $T_{q}$, extracted from the equilibrium excited-state population $P_1$ measured without applying any control pulses. The results are summarized in Fig. \ref{fig:FIG3}.  For each measurement, the number of readout shots is 100k--500k. A reset time of 5$\,$ms, about ten times more than the qubit relaxation time $T_1$, is used to ensure that the qubits fully relax to equilibrium before the next readout. The ground- and excited-state populations, $P_0$ and $P_1$, are extracted by fitting the readout histograms to a two-Gaussian model. Figure \ref{fig:FIG3}(a) shows a representative readout result from Q2. The extracted $P_1$ is as low as $0.02\%$, the initialization error without active reset. From the ratio $P_1/P_0$, the effective qubit temperature $T_{q}$ is obtained assuming a Boltzmann distribution. For the example shown in Fig. \ref{fig:FIG3}(a), the inferred effective qubit temperature is $T_{q}$ = 16.8$\,$mK. Similar results are observed for Q1 and Q3. In particular, Q3 exhibits an even lower $P_1 = 0.008\%$, corresponding to $T_{q}=17.1$$\,$mK, consistent with its slightly higher qubit frequency.

Figure \ref{fig:FIG3}(b) presents the extracted $T_{q}$ as a function of time for all three qubits. For comparison, the MXC temperature $T_{\rm{MX}}$, measured with a RuO$_2$ thermometer mounted on the MXC plate, is also plotted as the gray curve in Fig. \ref{fig:FIG3}(b). In contrast to the parity-switching rate, which continues to decrease over time without showing clear saturation, $T_q$ exhibits a much weaker temporal dependence. Specifically, $T_{q}$ decreases during the first few days after cooldown and then saturates at approximately $T_{\rm{MX}}$. This behavior indicates that the qubits are well thermalized to the phonon bath of the MXC plate. It also suggests that the qubit package and all electromagnetic shields are effectively thermalized to the same stage.

Figure \ref{fig:FIG3}(c) shows $T_{q}$ as a function of $T_{\rm{MX}}$ for all three qubits, with $T_{\rm{MX}}$ controlled via heating on Day~29. The dashed line indicates $T_{q} = T_{\rm{MX}}$. The measured data follow this line closely down to $T_{\rm{MX}}\approx$ 16.1$\,$mK with no clear sign of saturation in $T_{q}$. This result indicates that $P_1$ is primarily determined by the temperature of the phonon bath rather than by other mechanisms, such as nonequilibrium QPs. From the small discrepancy of about 2$\,$mK observed at $T_{\rm{MX}}$ = 16.1$\,$mK for Q2, and assuming that all excess excited-state population originates from nonequilibrium QPs, we can estimate an upper bound on the QP density \cite{jin2015thermal}. Under this assumption, the upper limit for the QP density is $x_{\rm{qp}} = 7.7 \times 10^{-9}$, which is consistent with and larger than the value extracted from $\Gamma_p^1$.

\begin{figure}
\centering
\includegraphics[width=\columnwidth]{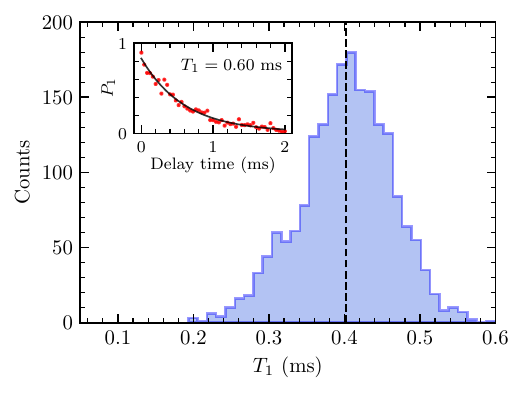}
\caption{Histogram of $T_1$ for Q2 on Day~24. The distribution has a mean of 402$\,$$\mu$s and a standard deviation of 60.7$\,$$\mu$s. The inset shows the longest measured $T_1$, reaching 600$\,$$\mu$s.}
\label{fig:FIG4}
\end{figure}

The energy relaxation time $T_{\rm{1}}$ of the qubits was systematically characterized. Using the active reset protocol, each $T_{\rm{1}}$ measurement required approximately 8$\,$s. To obtain sufficient statistics, 2000 repeated measurements were performed for each qubit. The averaged $T_{\rm{1}}$ values of different qubits range from 200 to 400$\,$$\mu$s. Figure \ref{fig:FIG4} shows a histogram of the measured $T_{\rm{1}}$ values for Q2. The distribution has an average value of 402$\,$$\mu$s with a standard deviation of 60.7$\,$$\mu$s. The longest $T_{\rm{1}}$ observed in this dataset reaches 600$\,$$\mu$s, as indicated in the inset.

In addition to these statistical measurements, the $T_{\rm{1}}$ values were monitored on a daily basis to examine their long-term stability (see Sec.~VII of the Supplemental Material~\cite{SM} for details). Aside from occasional sharp jumps and statistical fluctuations, no significant temporal trend in $T_{\rm{1}}$ was observed. The absence of a systematic time dependence, despite the decrease of $\Gamma_p$ over the same period, indicates that $T_1$ is not limited by QPs. After Day~15, QP tunneling contributes to relaxation at a rate below $\bar\Gamma_p^1\approx2.0$--$13$\,Hz (Sec.~IV.A of the Supplemental Material~\cite{SM}), at least two orders of magnitude below $1/T_1$ for all three qubits and about three for Q1 and Q2. The qubit relaxation is therefore primarily limited by material-related loss mechanisms. The observed sharp jumps are likely associated with two-level-system (TLS) defects, whose loss contributions and resonance frequencies can fluctuate due to low-frequency thermal fluctuators \cite{klimov2018fluctuations,schlor2019correlating}.

\begin{table*}[t]
\caption{Qubit coherence metrics in each shielding configuration, each measured in a single cooldown. Each value is averaged over repeated measurements, and $\Gamma_p$ is the lowest such value reached in each configuration. To streamline the discussion, Config.~C and cooldown A2 reported in an earlier version of this work are omitted. Also, cooldowns A3 and A1 in that version correspond to Configs.~A and F in the current version.}
\renewcommand{\arraystretch}{1.4}
\setlength{\tabcolsep}{6pt}
\resizebox{\textwidth}{!}{%
\begin{tabular}{|c|c|c|c|c|c|c|c|c|c|c|c|c|c|c|c|}
\hline
Configuration &
  \multicolumn{3}{c|}{\textbf{A}} &
  \multicolumn{3}{c|}{\textbf{B}} &
  \multicolumn{3}{c|}{\textbf{D}} &
  \multicolumn{3}{c|}{\textbf{E}} &
  \multicolumn{3}{c|}{\textbf{F}} \\ \hline
\makecell{In-line\\ HERD-2 filters} &
  \multicolumn{3}{c|}{Y} &
  \multicolumn{3}{c|}{Y} &
  \multicolumn{3}{c|}{N} &
  \multicolumn{3}{c|}{Y} &
  \multicolumn{3}{c|}{Y} \\ \hline
Indium seal ring &
  \multicolumn{3}{c|}{Y} &
  \multicolumn{3}{c|}{Y} &
  \multicolumn{3}{c|}{Y} &
  \multicolumn{3}{c|}{N} &
  \multicolumn{3}{c|}{Y} \\ \hline
\makecell{Number of\\ feedthroughs} &
  \multicolumn{3}{c|}{5} &
  \multicolumn{3}{c|}{16} &
  \multicolumn{3}{c|}{5} &
  \multicolumn{3}{c|}{5} &
  \multicolumn{3}{c|}{16} \\ \hline
\makecell{IR filters on\\ unused feedthroughs} &
  \multicolumn{3}{c|}{Y} &
  \multicolumn{3}{c|}{N} &
  \multicolumn{3}{c|}{Y} &
  \multicolumn{3}{c|}{Y} &
  \multicolumn{3}{c|}{N} \\ \hline
MXC shield &
  \multicolumn{3}{c|}{Y} &
  \multicolumn{3}{c|}{N} &
  \multicolumn{3}{c|}{Y} &
  \multicolumn{3}{c|}{Y} &
  \multicolumn{3}{c|}{Y} \\ \hline
\makecell{Cooldown duration\\ (days)}  &
  \multicolumn{3}{c|}{36} &
  \multicolumn{3}{c|}{9} &
  \multicolumn{3}{c|}{20} &
  \multicolumn{3}{c|}{19} &
  \multicolumn{3}{c|}{6} \\ \hline
Qubit &
  Q1 & Q2 & Q3 &
  Q1 & Q2 & Q3 &
  Q1 & Q2 & Q3 &
  Q1 & Q2 & Q3 &
  Q1 & Q2 & Q3 \\ \hline
$T_1\ (\mu \mathrm{s})$ &
  310 & 402 & 203 &
  247 & 276 & 194 &
  304 & 332 & 201 &
  296 & 295 & 197 &
  215 & 354 & 202 \\ \hline
$T_2\ (\mu \mathrm{s})$ &
  204 & 266 & 130 &
  152 & 177 & 117 &
   99 &  95 & 163 &
   87 &  95 & 105 &
  192 & 242 & 143 \\ \hline
$P_1\ (\%)$ &
  0.039 & 0.021 & 0.008 &
  0.078 & 0.045 & 0.061 &
  0.030 & 0.011 & 0.082 &
  0.42 & 0.19 & 0.21 &
  0.051 & 0.044 & 0.104 \\ \hline
$T_q\ (\mathrm{mK})$ &
  18.8 & 16.7 & 17.1 &
  20.7 & 18.4 & 21.8 &
  18.2 & 15.5 & 22.7 &
  27.0 & 22.5 & 26.2 &
  19.5 & 18.3 & 23.5 \\ \hline
$T_{\mathrm{MX}}\ (\mathrm{mK})$ &
  \multicolumn{3}{c|}{16--17} &
  \multicolumn{3}{c|}{11--13} &
  \multicolumn{3}{c|}{9--10} &
  \multicolumn{3}{c|}{15--16} &
  \multicolumn{3}{c|}{17--18} \\ \hline
$\Gamma_p\ (\mathrm{Hz})$ &
  0.074 & 0.069 & 0.111 &
  16.2 & 24.2 & 36.9 &
  0.396 & 0.262 & 0.560 &
  0.150 & 0.145 & 0.457 &
  0.8 & 1.1 & 4.8 \\ \hline
\end{tabular}%
}
\label{table:TABLE2}
\end{table*}

\section{Contribution of each shielding element}\label{sec:config}

To identify the role of each element, we compare configurations that differ in a single element (Table~\ref{table:TABLE2}). Relative to Config.~A, all four in-line HERD-2 filters are removed in Config.~D, the indium seal ring of the aluminum enclosure in Config.~E, and Config.~F uses a package with 16 SMA feedthroughs instead of 5. Configuration~B differs from Config.~F by the removal of the MXC shield. In all configurations, only the input and output lines are connected, and the unused feedthroughs are closed outside the package with SMA terminations or shorting caps. In the package with 5 feedthroughs, the three unused ones are also fitted with IR filters inside the package (Sec.~I of the Supplemental Material~\cite{SM}). At the same cooldown time, the rates in Config.~F exceed those in Config.~A by a factor of $2$--$7$, likely because of radiation entering through the 14 unused feedthroughs without IR filters. In the following, we compare Configs.~B, D and E with Config.~A, of which only Config.~B differs in the number of feedthroughs. Because $\Gamma_p$ decreases over weeks after cooldown, as shown in Fig.~\ref{fig:FIG2}(b), we compare the lowest rate reached in each configuration. This rate is the stationary value in Configs.~B, D and E. In Configs.~D and E, $\Gamma_p$ saturates, and in Config.~B, the rate on Day~9 exceeds that of Config.~A on the same day by more than an order of magnitude. In Config.~A, $\Gamma_p$ is still decreasing after 36 days, so that its lowest rate is an upper bound on this value.

\begin{figure}
\centering
\includegraphics[width=\columnwidth]{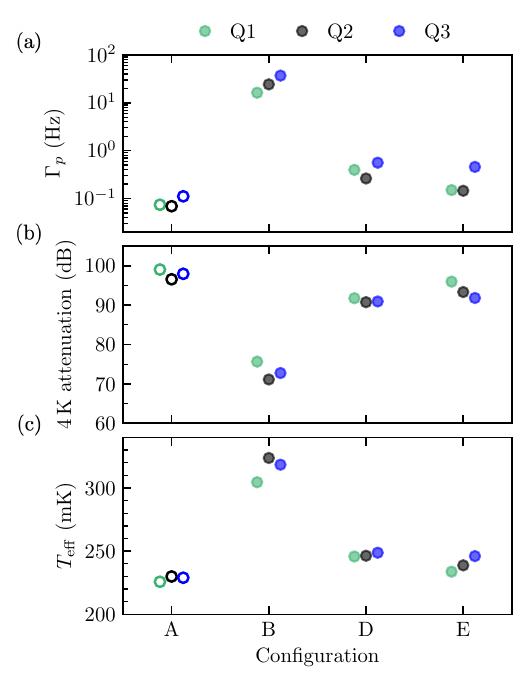}
\caption{Parity-switching rate, attenuation of 4\,K radiation, and effective blackbody temperature in each shielding configuration. (a) Lowest $\Gamma_p$ reached in each configuration for Q1--Q3. (b) Attenuation of 4\,K radiation, $\Gamma_p^{\mathrm{PAPS}}(4\,\mathrm{K})/\Gamma_p$, and (c) $T_{\mathrm{eff}}$, both obtained from Eq.~(\ref{eq:paps}) with the coupling efficiency of each qubit. Open symbols mark Config.~A, in which $\Gamma_p$ was still decreasing, so that its $T_{\mathrm{eff}}$ is an upper bound and its attenuation a lower bound.}
\label{fig:FIG5}
\end{figure}

To express the shielding performance in terms of the radiation reaching the device, we treat each qubit as an antenna coupled to free space, whose coupling efficiency $e_c(\omega)$ is the fraction of the incident radiation power transferred to the junction, obtained from electromagnetic simulations of its pads and junction (Sec.~II of the Supplemental Material~\cite{SM}). For blackbody radiation at temperature $T_{\mathrm{rad}}$, the PAPS rate is~\cite{liu2024quasiparticle}
\begin{equation}
\Gamma_p^{\mathrm{PAPS}}(T_{\mathrm{rad}})=\int_{(2\Delta+\delta\Delta)/\hbar}^{\infty} e_c(\omega)\,\bar n(\omega,T_{\mathrm{rad}})\,\frac{d\omega}{2\pi},
\label{eq:paps}
\end{equation}
where $\bar n(\omega,T_{\mathrm{rad}})=[\exp(\hbar\omega/k_BT_{\mathrm{rad}})-1]^{-1}$ is the mean photon occupation at frequency $\omega$, and the threshold $2\Delta+\delta\Delta$ is the sum of the gaps of the two electrodes, since pair breaking leaves one QP in each. Setting $\Gamma_p^{\mathrm{PAPS}}(T_{\mathrm{eff}})=\Gamma_p$ defines an effective blackbody temperature $T_{\mathrm{eff}}$~\cite{liu2024quasiparticle,Iaia2022PhononDT,ChargeParityRates}. Since $\Gamma_p$ contains every parity-switching mechanism, $T_{\mathrm{eff}}$ is an upper bound on the temperature of the radiation reaching the device. In Config.~A, the three qubits yield $T_{\mathrm{eff}}=226$--$230$\,mK. Equivalently, the attenuation of 4\,K radiation by the shielding and filtering, $\Gamma_p^{\mathrm{PAPS}}(4\,\mathrm{K})/\Gamma_p$, is $96$--$99$\,dB, a lower bound for the same reason. We take 4\,K as the reference because the thermal energy at this temperature is close to the pair-breaking energy, so that 4\,K radiation populates the entire pair-breaking band, and because all the shielding elements studied here are located below the 4\,K stage. For each qubit, a different reference shifts the attenuation of every configuration by the same amount and leaves their differences unchanged. These values of $T_{\mathrm{eff}}$ are below the $280$--$490$\,mK inferred with the same analysis in previous experiments~\cite{Iaia2022PhononDT,liu2024quasiparticle,ChargeParityRates}.    Figures~\ref{fig:FIG5}(a)--\ref{fig:FIG5}(c) show $\Gamma_p$, the attenuation of 4\,K radiation, and $T_{\mathrm{eff}}$ for Configs.~A, B, D and E. Configuration~F, measured for only 6 days, is not included. The decrease of the attenuation in Fig.~\ref{fig:FIG5}(b) when an element is removed quantifies the contribution of that element.

As shown in Fig.~\ref{fig:FIG5}, the MXC shield provides the largest suppression. Although Config.~B uses the package with 16 feedthroughs, the comparison between Configs.~F and A shows that the additional feedthroughs change $\Gamma_p$ by at most a factor of $2$--$7$, much less than the effect of removing the shield. We therefore compare Config.~B directly with Config.~A. Without the MXC shield, $\Gamma_p$ in Config.~B reaches $16$--$37$\,Hz, more than two orders of magnitude above the level reached in Config.~A (Fig.~\ref{fig:FIG5}(a)), whereas $T_{\mathrm{eff}}$ rises only to $304$--$324$\,mK (Fig.~\ref{fig:FIG5}(c)), reflecting its logarithmic dependence on $\Gamma_p$.  In Config.~B, the package is exposed to radiation from the warmer stages of the cryostat. Although the feedthroughs are sealed to the top plate with indium, the joints of the SMA connectors are not light-tight at the pair-breaking frequencies. Radiation can therefore enter the package through the 14 unused feedthroughs, which carry no IR filters. On the connected lines, it can also enter the coaxial segment between the outer and inner filters at a connector and leak out at the adapter before the inner filter, bypassing both filters. Imperfect seals of the enclosure provide another possible path, which Config.~E tests.

The in-line HERD-2 filters remove the contribution of the coaxial lines. Without them, $\Gamma_p$ in Config.~D remains at $0.26$--$0.56$\,Hz from Day~8 until the end of the 20-day cooldown (Sec.~V of the Supplemental Material~\cite{SM}), whereas in Config.~A it decreases by a factor of about two over the same period. A rate that no longer decreases indicates a stationary source, consistent with pair-breaking photons propagating along the unfiltered lines. The saturated rates exceed the lowest rates of Config.~A by a factor of about $4$--$5$. The attenuation is $6$--$7$\,dB lower than in Config.~A, and $T_{\mathrm{eff}}$ is $246$--$249$\,mK (Figs.~\ref{fig:FIG5}(b) and \ref{fig:FIG5}(c)). With the filters in place, the pair-breaking photons propagating along the lines are attenuated by more than $65$\,dB over $70$--$120$\,GHz, the band that carries more than $98\%$ of the pair-breaking photon flux at $T_{\mathrm{eff}}$ (Sec.~II of the Supplemental Material~\cite{SM}). The highest rate in Config.~D, $0.56$\,Hz, bounds the contribution of the two unfiltered lines. Assuming that the attenuation does not decrease above $120$\,GHz, the contribution of the two filtered lines is therefore below $0.56\,\mathrm{Hz}\times10^{-65/10}\approx2\times10^{-7}$\,Hz, five orders of magnitude below the level reached in Config.~A.

These results also set requirements for scaling up quantum processors, since the number of coaxial lines entering a package grows with the number of qubits. Because removing the filters from only two lines raised $\Gamma_p$ of all three qubits, the radiation entering through each line causes quasiparticle poisoning across the whole chip. Assuming that this contribution grows linearly with the number of lines, each line without filters adds $0.10$--$0.22$\,Hz to $\Gamma_p$. Starting from the rates measured in Config.~A, $\Gamma_p$ would then reach $1$\,Hz with at most $4$--$10$ unfiltered lines, whereas a processor with $100$ qubits and a few control lines per qubit requires several hundred lines. With the filters in place, each line contributes less than $10^{-7}$\,Hz, so that of order $10^{7}$ filtered lines would be required to reach $1$\,Hz. Filtering every line therefore removes the number of lines as a constraint on the parity-switching rate, provided that the feedthroughs and connectors are light-tight.

The indium seal makes the enclosure light-tight against the radiation that still reaches the region inside the MXC shield. Without it, $\Gamma_p$ in Config.~E saturates at $0.145$--$0.457$\,Hz, as in Config.~D, indicating a stationary source, consistent with radiation entering through the unsealed joint between the aluminum enclosure and the top plate. The saturated rates exceed the lowest rates of Config.~A by a factor of $2$--$4$ but remain below those of Config.~D. These rates correspond to $T_{\mathrm{eff}}=234$--$246$\,mK and an attenuation $3$--$6$\,dB below that of Config.~A. Removing the seal also raises $T_q$, as discussed below.

In contrast to $\Gamma_p$, $T_q$ is relatively insensitive to these changes, except for the removal of the indium seal. Across all configurations except Config.~E, $T_q$ lies between $15$ and $24$\,mK, whereas $\Gamma_p$ varies by more than two orders of magnitude. In Configs.~B and D, $T_{\mathrm{MX}}$ is $9$--$13$\,mK, below the lowest value reached in Config.~A, whereas $T_q$ exceeds $T_{\mathrm{MX}}$ by $5$--$14$\,mK, so that in these configurations $T_q$ does not follow $T_{\mathrm{MX}}$ (Sec.~\ref{sec:Tq}). The residual $P_1$ in these configurations therefore arises from excitation sources other than the thermal environment at $T_{\mathrm{MX}}$.   In Config.~E, $T_{\mathrm{MX}}$ is $15$--$16$\,mK, comparable to that in Config.~A, but $T_q$ rises to $22.5$--$27.0$\,mK, $6$--$9$\,mK above that in Config.~A. Since every QP tunneling event and every pair-breaking event at the junction switches the parity, the excitation rate due to QPs or pair-breaking photons is bounded by $\Gamma_p^0\approx\Gamma_p<0.5$\,Hz, far below the excitation rate $\Gamma_{\uparrow}=P_1/T_1\approx6$--$14$\,Hz implied by this population, so that neither process accounts for it. The excess population therefore arises either from radiation below the pair-breaking threshold or from a local phonon bath hotter than $T_{\mathrm{MX}}$, which would also raise $\Gamma_p$ through thermal activation of the resident QPs across $\delta\Delta$. The relaxation time $T_1$ remains between $190$ and $400\,\mu$s in all configurations, so that even the elevated parity-switching rates of Config.~B do not measurably limit $T_1$. In contrast, $T_2$ of Q1 and Q2 drops below $100\,\mu$s in Configs.~D and E, compared with $150$--$270\,\mu$s in the other configurations, indicating additional dephasing whose origin we have not identified.

These results suggest that, in Config.~A, free-space radiation primarily from the still shield is suppressed by two barriers in series. The outer barrier is the MXC shield, which attenuates the radiation reaching the inner barrier: the dedicated multi-layer shielding assembly (the yellow box in Fig.~\ref{fig:FIG1}(a)), whose indium seal ensures a light-tight enclosure and whose coaxial feedthroughs are filtered to block most of the residual radiation. Introducing an imperfection in the inner barrier alone, either removing the indium seal from the enclosure, as in Config.~E, or using unfiltered feedthroughs, as in Configs.~D and F, increases $\Gamma_p$ by only a factor of $2$--$7$. This relatively modest increase indicates that the MXC shield keeps the incident radiation sufficiently weak, even when the inner barrier is compromised. By contrast, removing the MXC shield, as in Config.~B, exposes the unfiltered feedthroughs directly to radiation from the warmer cryostat stages, causing $\Gamma_p$ to increase by more than two orders of magnitude relative to Config.~A. These comparisons suggest design rules for the packages of large-scale quantum processors. The enclosure should be light-tight, and the number of lines that require filtering depends on the target $\Gamma_p$, with each unfiltered line adding $0.10$--$0.22$\,Hz. Since unfiltered feedthroughs and connector joints are likely entry paths in Configs.~B and F, unused feedthroughs should be removed or filtered, and the number of connector joints between the filters should be minimized. In addition, the interior surfaces should be coated with an absorber~\cite{barends2011minimizing,kreikebaum2016optimization}.

\section{Conclusion}
We have suppressed the IR radiation reaching superconducting qubits to the point where it no longer limits quasiparticle poisoning. In the best configuration, the QP density reaches $x_{\mathrm{qp}}\approx4.11\times10^{-11}$, the lowest value reported to date. At this density, the QP-induced relaxation rate is more than three orders of magnitude below the measured relaxation rate, so that the resident QPs no longer limit qubit coherence, a prerequisite for fault-tolerant operation of superconducting quantum processors. The effective qubit temperature follows the MXC temperature down to about $16$\,mK and reaches $16.7$\,mK, corresponding to an initialization error of about $0.02\%$ for qubits near $3$\,GHz. The effective blackbody temperature of the radiation reaching the device is bounded to below $230$\,mK, equivalent to an attenuation of 4\,K radiation of at least $96$\,dB. Since it accounts for the antenna coupling of each qubit, this bound provides a metric to compare shielding performance across experiments.

Removing individual shielding elements quantifies their roles. The MXC shield provides the dominant suppression, the indium seal further lowers the parity-switching rate and the effective qubit temperature, and the in-line IR filters remove the contribution of the coaxial lines. Without these filters, a few lines would raise the parity-switching rate to 1\,Hz, whereas with them, of order $10^7$ lines would be required, so that the number of lines no longer limits the rate. The comparisons are consistent with the MXC shield and the enclosure acting as two barriers in series. With the MXC shield in place, a single imperfection of the enclosure raises the parity-switching rate by only a factor of $2$--$7$. These results translate into design rules for the packages of large-scale quantum processors, in which the enclosure is light-tight, unused feedthroughs are removed or filtered, and the number of filtered lines is set by the target parity-switching rate.

With all elements in place, the parity-switching rate continues to decrease over weeks after cooldown, indicating that it is dominated by a time-dependent source rather than by a stationary radiation background. Possible sources include stress relaxation in the package and slowly cooling components. Identifying and eliminating this source is the next step toward further reducing quasiparticle poisoning in superconducting quantum processors.

\section*{Data availability}
The data that support the findings of this study are available from the corresponding author upon reasonable request.

\begin{acknowledgments}
We thank Robert McDermott and Britton Plourde for fruitful discussion. We acknowledge support from the QC-Test at the Research Center for Critical Issues (RCCI), Academia Sinica, Taiwan.  This work was supported by the National Science and Technology Council (NSTC) in Taiwan via Grants No. NSTC 113-2119-M-001-008, NSTC 114-2119-M-001-004, and from Academia Sinica via Grants No. AS-GCP-112-M01, AS-GCS-114-M04, AS-KPQ-111-TQRB.

\end{acknowledgments}


\bibliography{apssamp}

\end{document}


\renewcommand{\thefigure}{S\arabic{figure}}
\setcounter{figure}{0}
\renewcommand{\theequation}{S\arabic{equation}}
\setcounter{equation}{0}
\renewcommand{\thetable}{S\arabic{table}}
\setcounter{table}{0}

\preprint{APS/123-QED}

\title{Supplemental Material for ``Suppression of Quasiparticle Poisoning to $10^{-11}$ Levels in Superconducting Qubits via Infrared Shielding''}

\author{Wei-En Lin}
\affiliation{Research Center for Critical Issues, Academia Sinica, Tainan, 711010, Taiwan}
\affiliation{Department of Physics, National Central University, Taoyuan, 320317, Taiwan}

\author{Chen-Hsun Ma}
\affiliation{Research Center for Critical Issues, Academia Sinica, Tainan, 711010, Taiwan}
\affiliation{Department of Physics, National Taiwan University, Taipei, 106319, Taiwan}

\author{Li-Chieh Hsiao}
\affiliation{Research Center for Critical Issues, Academia Sinica, Tainan, 711010, Taiwan}
\affiliation{Department of Physics, National Cheng Kung University, Tainan, 701401, Taiwan}

\author{Erh-Hsiang Yeh}
\affiliation{Department of Physics, National Central University, Taoyuan, 320317, Taiwan}

\author{Wei-Lun Peng}
\affiliation{Department of Physics, National Changhua University of Education, Changhua, 500207, Taiwan}
 
\author{Yu-Sen Wei}
\affiliation{Research Center for Critical Issues, Academia Sinica, Tainan, 711010, Taiwan}
\affiliation{Department of Physics, National Central University, Taoyuan, 320317, Taiwan}

\author{Hsi-Sheng Goan}
\affiliation{Department of Physics, National Taiwan University, Taipei, 106319, Taiwan}
\affiliation{Center for Quantum Science and Engineering, National Taiwan University, Taipei, 106319, Taiwan}
\affiliation{Physics Division, National Center for Theoretical Sciences, Taipei, 106319, Taiwan}

\author{Cen-Shawn Wu}
\affiliation{Research Center for Critical Issues, Academia Sinica, Tainan, 711010, Taiwan}
\affiliation{Department of Physics, National Changhua University of Education, Changhua, 500207, Taiwan}

\author{Chung-Ting Ke}\email{ctke@gate.sinica.edu.tw}
\affiliation{Research Center for Critical Issues, Academia Sinica, Tainan, 711010, Taiwan}
\affiliation{Institute of Physics, Academia Sinica, Taipei, 115201, Taiwan}

\author{Yung-Fu Chen}\email{yfuchen@cc.ncu.edu.tw}
\affiliation{Research Center for Critical Issues, Academia Sinica, Tainan, 711010, Taiwan}
\affiliation{Department of Physics, National Central University, Taoyuan, 320317, Taiwan}
\affiliation{Center of High Energy and High Field Physics, National Central University, Taoyuan, 320317, Taiwan}
\affiliation{Quantum Technology Center, National Central University, Taoyuan, 320317, Taiwan}
\affiliation{Taiwan Semiconductor Research Institute, National Institutes of Applied Research, Hsinchu, 300091, Taiwan}
\affiliation{National Center for Excellence in Quantum Information Science and Engineering, National Tsing Hua University, Hsinchu, 300044, Taiwan }

\author{Chii-Dong Chen}
\affiliation{Research Center for Critical Issues, Academia Sinica, Tainan, 711010, Taiwan}
\affiliation{Institute of Physics, Academia Sinica, Taipei, 115201, Taiwan}

\maketitle
\onecolumngrid

\addtocontents{toc}{\protect\setcounter{tocdepth}{2}} 
\tableofcontents 
\newpage

\section{Experimental setup}\label{sec:Experimental setup}

\begin{figure} 
\centering
\includegraphics[width=0.8\columnwidth]{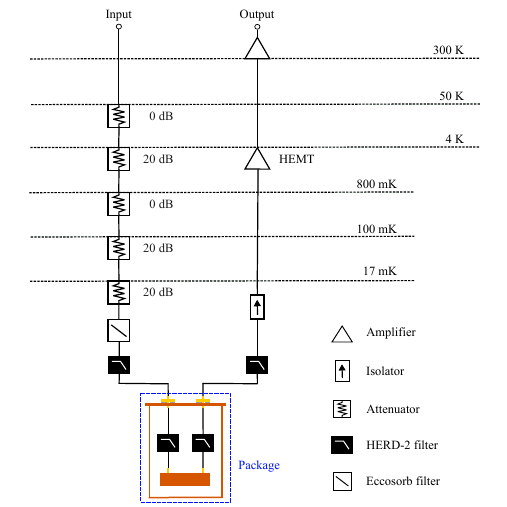} 
\caption{Schematic of the experimental wiring and measurement setup. Attenuation values are given for each stage of the input line. Two HERD-2 filters are mounted inside the package, one on each of the input and output lines, in addition to those outside. Details of the package are shown in Fig.~1 of the main text.}
\label{fig:S1}
\end{figure}

Figure~\ref{fig:S1} shows the wiring of the measurement chain. The input drive line is attenuated by 20~dB at each of the 4~K, 100~mK and mixing-chamber (MXC) stages, giving 60~dB in total. To suppress stray infrared (IR) radiation propagating along the coaxial lines, filters are placed on both signal paths. The input line carries an absorptive Eccosorb filter (Bluefors, low-loss bulkhead IR filter) followed by a non-magnetic High-Energy Radiation-Drain filter (Sweden Quantum, HERD-2) outside the package, and a second HERD-2 filter inside it. The output line carries one HERD-2 filter inside the package and one outside, followed by an isolator at the MXC stage and a HEMT amplifier at 4~K. No Eccosorb filter is placed on the output line, since its insertion loss would degrade the noise temperature of the amplification chain. The HERD-2 filters provide the IR filtering on this path instead, with a passband insertion loss below 0.13~dB over DC--13.5~GHz and more than 65~dB of attenuation over 70--120~GHz. The relevance of this band to the pair-breaking photon flux is quantified in Sec.~\ref{sec:antenna}. The filters outside the package are thermally anchored to the MXC plate, and those inside the package to the copper top plate of the package. Unused feedthroughs of the package are closed with SMA terminations or shorting caps to block radiation entering through the connectors. In the package with 5 feedthroughs, used in Configs.~A, D and E, the three unused feedthroughs are additionally fitted with IR filters (Quantum Microwave, QMC-CRYOIRF-003MF-S) inside the package, whereas the 14 unused feedthroughs of the package with 16 feedthroughs, used in Configs.~B and F, carry no IR filters.

The same wiring is used in all configurations except Config.~D, in which all four HERD-2 filters are removed. Relative to Config.~A, the indium seal ring of the aluminum enclosure is removed in Config.~E, and Configs.~B and F use the package with 16 feedthroughs, with the MXC shield additionally removed in Config.~B (Table~II of the main text).

\section{Antenna mode simulation}\label{sec:antenna}

Following the framework developed in Ref.~\cite{liu2024quasiparticle}, we evaluate the coupling efficiency to IR radiation $e_c$ by modeling the floating transmon as an antenna coupled to free space. The coupling efficiency is given as
\begin{equation}
    e_c(\omega) = 1 - \left| \frac{Z_{\rm{rad}}(\omega) - Z_j^*(\omega)}{Z_{\rm{rad}}(\omega) + Z_j(\omega)} \right|^2,
\end{equation}
where $Z_{\rm{rad}}$ is the frequency-dependent radiation impedance associated with the capacitor-pad structure, and $Z_{j}$ is the junction impedance. Physically, $e_c$ characterizes how efficiently incident electromagnetic radiation can transfer power into the junction degree of freedom through impedance matching.

\begin{figure}
\centering
\includegraphics[width=\columnwidth]{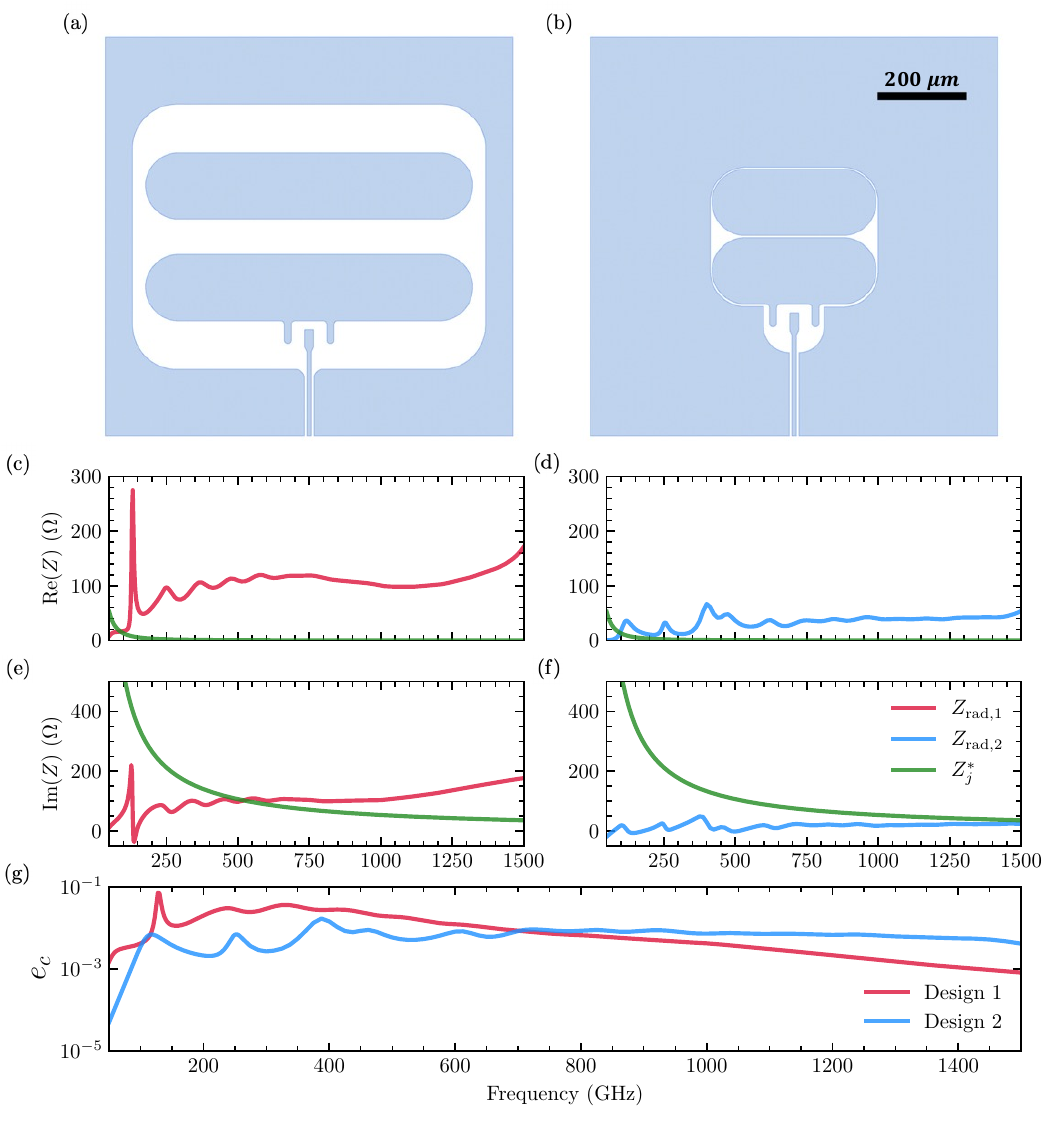}
\caption{Numerical analysis of the antenna-coupling model and impedance matching. (a) Simulated geometry of \textbf{Design 1} (based on Q1). (b) Simulated geometry of \textbf{Design 2}, featuring capacitor pads with half the area of \textbf{Design 1} and narrower gaps for comparison. (c)--(f) Frequency-dependent real and imaginary components of the radiation impedance $Z_{\rm{rad},i}$ (where $i=1, 2$ denotes Design 1 and 2, respectively), calculated using \texttt{openEMS} and plotted alongside the modeled complex conjugate of the junction impedance, $Z_j^*$. (g) Calculated coupling efficiency $e_c$ as a function of frequency. The comparison illustrates that \textbf{Design 1} couples more strongly than \textbf{Design 2} below about $700$\,GHz.}
\label{fig:S2}
\end{figure}

To analyze the role of device geometry, we take the geometry of Q1 (denoted here as \textbf{Design 1}) as a representative example, as shown in Fig.~\ref{fig:S2}(a). The radiation impedance $Z_{\rm{rad}}$ is calculated using the open-source electromagnetic solver \texttt{openEMS} \cite{openEMS}. For comparison, we also consider a reference geometry (\textbf{Design 2}) with the same charging energy $E_C$, but with capacitor pads having approximately half the area of \textbf{Design 1} and correspondingly reduced gap dimensions, as illustrated in Fig.~\ref{fig:S2}(b). The calculated real and imaginary parts of $Z_{\rm{rad}}$ are shown in Figs.~\ref{fig:S2}(c) and \ref{fig:S2}(e) for \textbf{Design 1} and in Figs.~\ref{fig:S2}(d) and \ref{fig:S2}(f) for \textbf{Design 2}.

The two designs exhibit qualitatively different electromagnetic behavior. Due to its larger pads and greater separation from the surrounding ground plane, \textbf{Design 1} behaves more like a distributed antenna structure. The weaker screening by the ground plane allows stronger fringing electromagnetic fields and more efficient coupling to free space, leading to a substantially larger real part of the radiation impedance compared to \textbf{Design 2}. In contrast, the more compact geometry of \textbf{Design 2} confines the electric fields more strongly near the substrate and the ground, suppressing radiation into free space. The imaginary part of the radiation impedance further reveals the distinct reactive characteristics of the two geometries. Apart from a sharp resonance of \textbf{Design 1} near $133$\,GHz, the imaginary parts of both designs remain mostly positive above approximately $100$\,GHz, indicating predominantly inductive behavior. Because \textbf{Design 1} has longer pads, its effective inductive response is stronger, resulting in a larger imaginary part of the impedance than \textbf{Design 2} over the relevant frequency range. 

The junction impedance is modeled as a parallel $RC$ circuit,
\begin{equation}
    Z_j(\omega) = \frac{R_n}{1 + j\omega R_n C_j},
\end{equation}
where we use the normal-state resistance of Q1, $R_n = 21.4\,\text{k}\Omega$, and a junction capacitance $C_j = 3\,\text{fF}$. Its complex conjugate $Z_j^*$ is also plotted in Figs.~\ref{fig:S2}(c)--(f). Maximum power transfer occurs when the conjugate matching condition, $Z_{\rm{rad}}=Z_j^*$, is approximately satisfied. Figure~\ref{fig:S2}(g) presents the calculated coupling efficiency $e_c(\omega)$ for both designs. Due to its larger inductive reactance, \textbf{Design 1} provides improved impedance matching to the junction over the 100--400$\,$GHz frequency range, yielding $e_c > 0.01$ throughout much of this band. In contrast, \textbf{Design 2} couples most strongly near $390$\,GHz and exceeds \textbf{Design 1} only above about $700$\,GHz. \textbf{Design 1}, the pad geometry of Q1 and Q3, thus couples efficiently at lower frequencies than more compact designs such as \textbf{Design 2}.

\begin{figure}[t]
\centering
\includegraphics[width=\columnwidth]{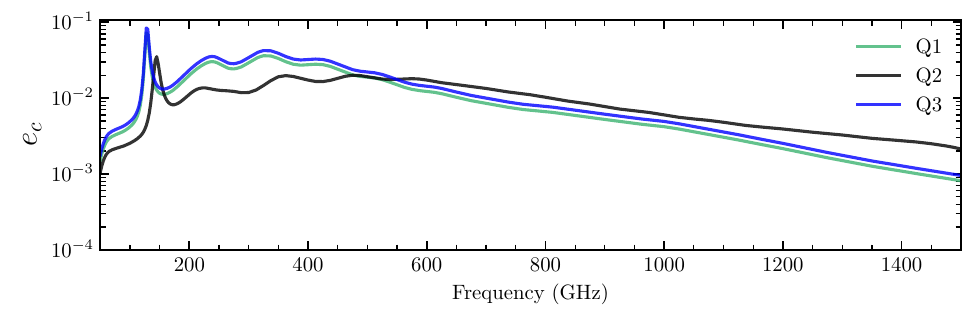}
\caption{Coupling efficiency $e_c$ of Q1--Q3, calculated from the simulated radiation impedance and the junction impedance of each qubit. Q1 and Q3 share the same pad geometry and differ only in $R_n$.}
\label{fig:S3}
\end{figure}

The simulated $e_c(\omega)$ can be combined with the measured parity-switching rate to place an upper limit on the IR radiation reaching the device. In photon-assisted parity switching (PAPS), a photon with energy above the pair-breaking threshold is absorbed at the junction, breaking a Cooper pair and depositing one quasiparticle (QP) on each electrode, so that the charge parity is switched without any pre-existing QP~\cite{houzet2019,diamond2022distinguishing}. Taking the measured critical temperature of our aluminum films, $T_c=1.23$~K, the BCS relation $\Delta=1.76\,k_BT_c$ gives $\Delta=186.5\,\mu\mathrm{eV}$. Treating the pads as a single-mode antenna looking at blackbody radiation at temperature $T_{\mathrm{rad}}$, the mean photon occupation of the mode is the Bose--Einstein factor $\bar n(\omega,T_{\mathrm{rad}})=[\exp(\hbar\omega/k_BT_{\mathrm{rad}})-1]^{-1}$, and the PAPS rate can be calculated as
\begin{equation}
\Gamma_p^{\mathrm{PAPS}}(T_{\mathrm{rad}})=\int_{(2\Delta+\delta\Delta)/\hbar}^{\infty} e_c(\omega)\,\bar n(\omega,T_{\mathrm{rad}})\,\frac{d\omega}{2\pi},
\label{eq:paps}
\end{equation}
where $\delta\Delta$ is the gap difference between the two electrodes and the threshold is set by the sum of the two gaps, since pair breaking at the junction leaves one QP in each electrode. With $\delta\Delta/h=0.71$--$1.19$\,GHz from Sec.~\ref{sec:gap}, the threshold lies at $90.9$--$91.4$\,GHz. The measured switching rate $\Gamma_p$ contains every parity-switching mechanism, so the PAPS contribution can only be smaller. Setting $\Gamma_p^{\mathrm{PAPS}}(T_{\mathrm{rad}})=\Gamma_p$ and solving for $T_{\mathrm{rad}}$, as in Refs.~\cite{liu2024quasiparticle,Iaia2022PhononDT,ChargeParityRates}, therefore returns an upper bound on the effective blackbody temperature, $T_{\mathrm{eff}}$. Table~\ref{table:S1} lists $T_{\mathrm{eff}}$ for Config.~A, obtained from the simulated $e_c$ of each qubit (Fig.~\ref{fig:S3}) and its measured parity-switching rate. The three qubits yield upper bounds of $T_{\mathrm{eff}}=226$--$230$\,mK.

\begin{table}[h!]
\caption{Upper bounds on the effective blackbody temperature in Config.~A. $\Gamma_p^{4\mathrm{K}}\equiv\Gamma_p^{\mathrm{PAPS}}(4\,\mathrm{K})$ denotes the PAPS rate for a 4\,K blackbody, and $R_n$ is inferred from $E_J$ via the Ambegaokar--Baratoff relation \cite{ambegaokar1963}.}
\label{table:S1}
\begin{ruledtabular}
\begin{tabular}{lcccc}
 & $R_n$ (k$\Omega$) & $\Gamma_p$ (Hz) & $T_{\mathrm{eff}}$ (mK) & $\Gamma_p^{4\mathrm{K}}/\Gamma_p$ \\
\colrule
Q1 & 21.38 & 0.074 & 226 & $8.0\times10^{9}$ \\
Q2 & 21.05 & 0.069 & 230 & $4.5\times10^{9}$ \\
Q3 & 18.23 & 0.111 & 229 & $6.3\times10^{9}$ \\
\end{tabular}
\end{ruledtabular}
\end{table}

The last column of Table~\ref{table:S1} compares $\Gamma_p$ with $\Gamma_p^{4\mathrm{K}}$, the PAPS rate expected if the device were exposed to blackbody radiation at 4\,K. This ratio quantifies the combined suppression of 4\,K radiation by the shielding and filtering, and lies between $4.5\times10^{9}$ and $8.0\times10^{9}$ for the three qubits. It is a lower bound, for the same reason that $T_{\mathrm{eff}}$ is an upper bound. Since the thermal excited-state population is below $10^{-3}$, the measured rate closely approximates the ground-state parity-switching rate $\Gamma_p^0$. How close both bounds lie to the true values therefore depends on the PAPS fraction of $\Gamma_p^0$. Measuring $\Gamma_p^0$ as a function of temperature could in principle determine this fraction, since tunneling of resident QPs increases with temperature whereas PAPS is independent of the bath temperature~\cite{connolly2024coexistence}. The limitation of this approach for the small gap differences of our junctions and the PAPS contribution to the excited-state parity-switching rate $\Gamma_p^1$ are discussed in Sec.~\ref{sec:gap}. The decrease of $\Gamma_p$ over the cooldown points to a time-dependent source, such as phonon bursts from stress relaxation or thermal radiation from slowly cooling components (Sec.~III of the main text).

Only photons with energy within a few $k_BT_{\mathrm{eff}}$ above the threshold contribute significantly to the photon flux, so in-line filtering only has to be effective over this energy range. The fraction of the pair-breaking photon flux below a cut-off $\omega_c$ is given by
\begin{equation}
\eta(\omega_c)=\frac{\displaystyle\int_{(2\Delta+\delta\Delta)/\hbar}^{\omega_c} e_c(\omega)\,\bar n(\omega,T_{\mathrm{rad}})\,d\omega}
{\displaystyle\int_{(2\Delta+\delta\Delta)/\hbar}^{\infty} e_c(\omega)\,\bar n(\omega,T_{\mathrm{rad}})\,d\omega}.
\label{eq:fluxfrac}
\end{equation}
Although $e_c$ is not monotonic and carries structure well above the threshold, the exponential suppression of $\bar n$ dominates by many orders of magnitude, so $\eta$ is set by $e_c$ between the threshold and the cut-off. Evaluating Eq.~(\ref{eq:fluxfrac}) at $T_{\mathrm{rad}}=T_{\mathrm{eff}}$ gives $\eta=98.2$--$99.6\%$ for the three qubits at $\omega_c/2\pi=120$\,GHz, the upper edge of the band over which the HERD-2 attenuation is specified (Sec.~\ref{sec:Experimental setup}). The HERD-2 filters therefore block the frequencies that carry the pair-breaking photon flux, and their removal, described in Sec.~\ref{sec:P_CD}, tests the contribution of the coaxial lines.

\section{Hidden Markov Models}\label{sec:hmm}

To accurately characterize the parity-switching rate $\Gamma_p$, we employ a Hidden Markov Model (HMM) to mitigate the influence of parity mapping errors. The HMM provides a probabilistic framework for systems in which the intrinsic state dynamics are not directly observable but must instead be inferred from noisy measurements~\cite{rabiner1989hmm}. 

Within this framework, the actual parity state is treated as a hidden variable, while the experimentally assigned parity outcome obtained from mapping the parity onto the qubit state is regarded as a noisy observation. The hidden states evolve according to a Markov process, and each hidden state probabilistically generates a measurement outcome. The HMM therefore enables simultaneous estimation of model parameters and reconstruction of the most probable hidden-state trajectory from experimental data. In this work, these procedures are implemented using the \texttt{hmmlearn} package \cite{hmmlearn}, where the Viterbi algorithm is employed to reconstruct the most likely latent parity sequence.

To analyze the parity dynamics, we model the measured parity time trace using a two-state HMM. The hidden state at measurement step $t$ is denoted by $q_t \in \{e,o\}$, where $e$ and $o$ represent the even- and odd-parity states, respectively. Assuming that the parity dynamics follow a first-order Markov process, the parity state at the next time step depends only on the current state,
\begin{equation}
    P(q_{t+1} \mid q_t, q_{t-1}, \cdots) = P(q_{t+1} \mid q_t).
\end{equation}
The stochastic evolution of the hidden states is described by the transition matrix
\begin{equation}
    A =
    \begin{pmatrix}
        P(e \mid e) & P(o \mid e) \\
        P(e \mid o) & P(o \mid o)
    \end{pmatrix}.
\end{equation}
The time interval between two consecutive parity measurements is denoted as $\Delta t$. In the limit where the switching probability between consecutive measurements is small ($\Delta t \ll 1/\Gamma_p$), the parity-switching rates are determined from the off-diagonal transition probabilities,
\begin{equation}
    \Gamma_{e\rightarrow o} = \frac{P(o \mid e)}{\Delta t},
    \qquad
    \Gamma_{o\rightarrow e} = \frac{P(e \mid o)}{\Delta t},
\end{equation} and the average parity-switching rate can be defined as
\begin{equation}
    \Gamma_p=\frac{\Gamma_{e\rightarrow o}+\Gamma_{o\rightarrow e}}{2}.
\end{equation}

The observed parity outcome $x_t \in \{e,o\}$ is probabilistically related to the hidden state $q_t$ through the emission probability $\tilde{P}(x_t \mid q_t)$, which specifies the conditional probability of obtaining a particular measurement outcome given the underlying parity state $q_t$. The corresponding emission matrix is given by
\begin{equation}
    B =
    \begin{pmatrix}
        \tilde{P}(e \mid e) & \tilde{P}(o \mid e) \\
        \tilde{P}(e \mid o) & \tilde{P}(o \mid o)
    \end{pmatrix}.
\end{equation}
The off-diagonal elements of $B$ quantify parity mapping errors arising from imperfect parity-to-qubit mapping and finite qubit readout fidelity, where the average parity mapping fidelity is given by  
\begin{equation}
F= \frac{\tilde{P}(e \mid e)+ \tilde{P}(o \mid o)}{2}.
\end{equation}
By separately modeling the intrinsic parity dynamics through the transition matrix $A$ and the parity-state assignment through the emission matrix $B$, the HMM provides a robust probabilistic framework for reconstructing the most likely parity trajectory while mitigating the impact of parity mapping errors.

\begin{figure}
\centering
\includegraphics[width=\columnwidth]{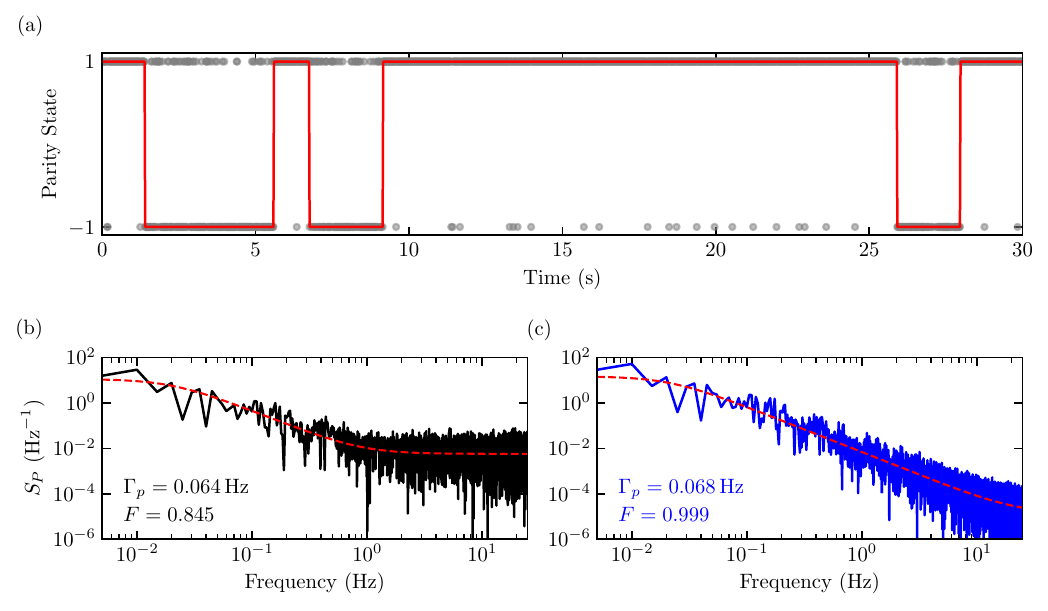}
\caption{Hidden Markov Model (HMM) reconstruction of parity dynamics. 
(a) A representative 30-s segment extracted from a 200-s sampling duration. Raw parity measurements and the HMM-reconstructed trajectory are depicted as gray dots and a solid red line, respectively. (b)(c) Power spectral densities (PSDs) obtained from the raw (b) and reconstructed (c) data. Dashed red lines indicate Lorentzian fits according to Eq.~\eqref{eq:PSD_model}.}
\label{fig:S4}
\end{figure}

The parity state was sampled every $\Delta t = 20\,$ms over a total duration of 200$\,$s using the Ramsey-based parity-mapping sequence described in Ref.~\cite{serniak2018hot}, with the conditional $\pi$ pulse applied only to the odd-parity state. In Fig.~\ref{fig:S4}(a), a representative 30-s segment of the measured parity time series of Q2 captured on Day~34 in Config.~A is displayed (gray dots). To mitigate parity mapping errors, the raw data are reconstructed into the most probable underlying parity trajectory using an HMM (red line). The HMM analysis yields the estimated transition and emission matrices 
\begin{equation}
    A \approx
    \begin{pmatrix}
        0.9971 & 0.0029 \\
        0.0019 & 0.9981
    \end{pmatrix},
    \qquad
    B \approx
    \begin{pmatrix}
        0.7654 & 0.2346 \\
        0.0294 & 0.9706
    \end{pmatrix},
\end{equation}
from which we obtain $\Gamma_p = 0.12\,$Hz and $F = 0.868$.  The conditional $\pi_o$ pulse generates a pronounced excited-state population contrast between parity manifolds. The resulting $T_1$-limited relaxation introduces the observed asymmetry in the emission matrix $B$. Both the raw and reconstructed time series are transformed into the frequency domain via a Fourier transform to determine the power spectral density (PSD). As illustrated in Fig.~\ref{fig:S4}(b) and (c), the switching rates $\Gamma_p$ are extracted by fitting the PSD to a Lorentzian model  \cite{serniak2018hot, riste2013millisecond}
\begin{equation}
\label{eq:PSD_model}
S_P(f) = F^2 \frac{4\Gamma_p}{(2\pi f)^2 + (2\Gamma_p)^2} + (1-F^2)\Delta t.
\end{equation}
HMM reconstruction improves the mapping fidelity $F$, increasing from $0.845$ to $0.999$. Given that a lower $F$ typically leads to an underestimation of $\Gamma_p$ in the Lorentzian fit, we observe a minor discrepancy from the raw $\Gamma_p$ of $0.064\,$Hz to the HMM-reconstructed value of $0.068\,$Hz.  Notably, the $\Gamma_p$ estimated from the transition matrix ($0.12\,$Hz) exceeds the value extracted from PSD fitting ($0.068\,$Hz) as the former is more susceptible to high-frequency spurious transitions. The PSD analysis further distinguishes the slow parity-switching dynamics from the nearly frequency-independent background arising from parity mapping errors. Additionally, the $F$ values obtained from the $B$ matrix ($0.868$) and the PSD analysis ($0.845$) remain consistent with each other. The PSD of the reconstructed trajectory exhibits a clear Lorentzian profile with a suppressed noise floor which validates that the HMM effectively preserves the intrinsic switching dynamics while filtering out parity mapping errors. Accordingly, this HMM-based filtering approach was applied to all parity data presented in the main text to mitigate the influence of mapping infidelities.

\section{Gate-charge dependence of parity-switching rate}\label{sec:gate_gamma_p}

The parity-switching rates $\Gamma_p^0$ and $\Gamma_p^1$ reported in Sec.~III of the main text show a dependence on the gate charge $n_g$. Before turning to extrinsic origins, we ask whether such a dependence can be generated by the QP tunneling rate itself. The usual treatment invokes the transmon limit $E_J\gg E_C$, with $E_J$ and $E_C$ the Josephson and charging energies~\cite{catelani2011relaxation,Glazman2021Bogoliubov}, in which the leading-order charge-parity matrix elements are independent of $n_g$, and this limit is adopted in recent analyses of parity switching~\cite{connolly2024coexistence,nho2026recovery}. The gate charge then enters only through corrections proportional to the exponentially small charge dispersion of the transmon levels~\cite{catelani2014parity,marchegiani2022asymmetric}. Here we compute the QP tunneling rate $\Gamma_{ij}$ for each qubit transition $|i\rangle\!\rightarrow\!|j\rangle$, with the charge-parity matrix elements evaluated between eigenstates of the Cooper-pair-box Hamiltonian, obtained exactly by numerical diagonalization in the charge basis.

A charge-parity switch occurs when a single QP tunnels across the Josephson junction. Because the gap of aluminum increases as the film is made thinner~\cite{Court_2008}, the two electrodes of our junction have different superconducting gaps. For our devices, the $100\;\mathrm{nm}$ top electrode is the low-gap side, with gap $\Delta$, and the $50\;\mathrm{nm}$ bottom electrode the high-gap side, with gap $\Delta+\delta\Delta$. A QP residing in the low-gap film must therefore acquire an energy $\delta\Delta$ in order to cross. In $|0\rangle$, the QP can acquire that energy only from the phonon bath, with which its energy distribution is thermalized~\cite{connolly2024coexistence}, so that $\Gamma_p^0$ is thermally activated. In $|1\rangle$, the qubit can instead supply its excitation energy $hf_{01}$ to the tunneling QP. For $\delta\Delta>hf_{01}$, the qubit energy lowers the activation energy of $\Gamma_p^1$ to $\delta\Delta-hf_{01}$. In the opposite case, $\delta\Delta<hf_{01}$, the qubit energy exceeds the gap difference and the QP crosses without thermal activation~\cite{nho2026recovery}. Therefore $\Gamma_p^1$ greatly exceeds the thermally activated $\Gamma_p^0$.

Following the Fermi-golden-rule treatment of Refs.~\cite{catelani2011relaxation,marchegiani2022asymmetric}, in the form given in Refs.~\cite{connolly2024coexistence,nho2026recovery}, the rate for a parity switch accompanied by the qubit transition $|i\rangle\!\rightarrow\!|j\rangle$ is
\begin{equation}
\Gamma_{ij}=\frac{16E_J}{h}
\left[S_-\bigl|\langle j,o|\cos\tfrac{\hat\varphi}{2}|i,e\rangle\bigr|^{2}
+S_+\bigl|\langle j,o|\sin\tfrac{\hat\varphi}{2}|i,e\rangle\bigr|^{2}\right],
\label{eq:fgr}
\end{equation}
where $\hat\varphi$ is the superconducting phase operator across the junction, and $e$ and $o$ denote the even and odd charge-parity states. The operators $\cos\frac{\hat\varphi}{2}$ and $\sin\frac{\hat\varphi}{2}$ describe the transfer of a single electron across the junction~\cite{catelani2014parity} and connect states of opposite parity. Since every $\Gamma_{ij}$ is a parity switch, the parity-switching rate with the qubit in $|i\rangle$ is the sum over all final states $j$, $\Gamma_p^0=\sum_j\Gamma_{0j}$ and $\Gamma_p^1=\sum_j\Gamma_{1j}$. The energy transferred to the tunneling QP is set by $hf_{ij}$, the energy absorbed by the qubit, with $f_{ij}\equiv f_j-f_i$. For relaxation, $f_{10}=-f_{01}$ is negative, and the qubit supplies energy to the QP as described above. The structure factors $S_\pm$ collect the BCS coherence factors, the densities of states and the QP occupation. Their dependence on the gap difference is conveniently expressed through the two activation exponents
\begin{equation}
\alpha_{ij}=\frac{\delta\Delta+hf_{ij}}{2k_BT},\qquad
\beta_{ij}=\frac{\delta\Delta-hf_{ij}}{2k_BT},
\label{eq:albe}
\end{equation}
which correspond to tunneling out of the low- and high-gap film, respectively. Here $T$ is the temperature of the phonon bath, taken equal to the measured MXC temperature $T_{\mathrm{MX}}$. The resident QPs are assumed to be thermalized to this bath despite their nonequilibrium density, as noted above.  The QPs occupy energies within a few $k_BT$ of the gap edge. Since $k_BT$, $\delta\Delta$ and $h|f_{ij}|$ are all much smaller than $\Delta$, we expand the exact structure factors of Ref.~\cite{connolly2024coexistence} about the gap edge and neglect corrections of relative order $k_BT/\Delta$, $\delta\Delta/\Delta$ and $h|f_{ij}|/\Delta$. Adding the contributions from both tunneling directions between the two films,
\begin{align}
S_+&=x_{\mathrm{qp}}\sqrt{\frac{\Delta}{2\pi k_BT}}\,e^{-\alpha_{ij}}
\left[K_0(|\alpha_{ij}|)+K_0(|\beta_{ij}|)\right],
\label{eq:Splus}\\[4pt]
S_-&=x_{\mathrm{qp}}\sqrt{\frac{\Delta}{2\pi k_BT}}\,
\frac{k_BT}{\bar\Delta}\,e^{-\alpha_{ij}}\left[|\alpha_{ij}|K_1(|\alpha_{ij}|)
+|\beta_{ij}|K_1(|\beta_{ij}|)\right],
\label{eq:Sminus}
\end{align}
with $K_0$ and $K_1$ the modified Bessel functions of the second kind and $\bar\Delta=\Delta+\delta\Delta/2$ the mean gap. Throughout, $x^{L}_{\mathrm{qp}}$ and $x^{H}_{\mathrm{qp}}$ denote the QP densities in the low- and high-gap films, normalized to the density of Cooper pairs, and $x_{\mathrm{qp}}\equiv x^{L}_{\mathrm{qp}}$ is the quantity reported in the main text. Since $x^{H}_{\mathrm{qp}}$ is smaller than $x^{L}_{\mathrm{qp}}$ by a factor of $7$--$28$ for our devices (Table~\ref{tab:gap}), $x_{\mathrm{qp}}$ is close to the total resident QP density. The two directions share the common factor $e^{-\alpha_{ij}}$ once the equilibrium ratio $x^{H}_{\mathrm{qp}}/x^{L}_{\mathrm{qp}}=\sqrt{\Delta/(\Delta+\delta\Delta)}\,e^{-\delta\Delta/k_BT}$ is inserted. Here the QPs on both sides of the junction share the energy distribution thermalized at $T$, so that their densities differ only by this equilibrium ratio.  In Eqs.~(\ref{eq:Splus})--(\ref{eq:Sminus}), the Bessel functions take $|\alpha_{ij}|$ and $|\beta_{ij}|$ as arguments. One of the two exponents becomes negative for $\delta\Delta<h|f_{ij}|$, and with these absolute values the expressions hold in this regime as well as for $\delta\Delta>h|f_{ij}|$~\cite{nho2026recovery}.

The gate charge enters Eqs.~(\ref{eq:albe})--(\ref{eq:Sminus}) only through the transition frequencies $f_{ij}$ in $\alpha_{ij}$ and $\beta_{ij}$. For the transitions that dominate the rates, the charge dispersion of the transmon levels makes these frequencies vary with $n_g$ by far less than $k_BT/h$, so $\alpha_{ij}$ and $\beta_{ij}$, and with them the structure factors, are effectively independent of $n_g$. Any intrinsic $n_g$ dependence must therefore come from the matrix elements, and we now evaluate them without transmon-limit approximation.

The Cooper-pair-box Hamiltonian in the charge basis reads~\cite{koch2007charge}
\begin{equation}
\hat H(n_g)=4E_C\sum_n (n-n_g)^2\,|n\rangle\langle n|
-\frac{E_J}{2}\sum_n\bigl(|n+1\rangle\langle n|+|n\rangle\langle n+1|\bigr),
\label{eq:H}
\end{equation}
where $|n\rangle$ is the state with $n$ Cooper pairs transferred across the junction and $n_g$ is the offset charge in units of $2e$. In this basis, $e^{\pm i\hat\varphi}$ shifts $n$ by $\pm1$. We write $\cos\frac{\hat\varphi}{2}$ and $\sin\frac{\hat\varphi}{2}$ in terms of $e^{\pm i\hat\varphi/2}$. These operators shift the charge by half a Cooper pair, $e^{\pm i\hat\varphi/2}|n\rangle=|n\pm\tfrac12\rangle$. The even- and odd-parity eigenstates are expanded over integer and half-integer charges, respectively,
\begin{equation}
|i,e\rangle=\sum_n c^{(i)}_n(n_g)\,|n\rangle,\qquad
|j,o\rangle=\sum_m c^{(j)}_m\bigl(n_g-\tfrac12\bigr)\,|m+\tfrac12\rangle,
\label{eq:expand}
\end{equation}
with real coefficients given by the eigenvectors of $\hat H(n_g)$ and $\hat H(n_g-\tfrac12)$, and $n$ and $m$ running over all integers. Acting on the even state gives $e^{\pm i\hat\varphi/2}|i,e\rangle=\sum_n c^{(i)}_n(n_g)\,|n\pm\tfrac12\rangle$. By Eq.~(\ref{eq:expand}), the odd state has coefficient $c^{(j)}_n$ at charge $n+\tfrac12$ and $c^{(j)}_{n-1}$ at charge $n-\tfrac12$, so projecting onto $|j,o\rangle$ yields
\begin{align}
M^+_{ji}&\equiv\langle j,o|e^{+i\hat\varphi/2}|i,e\rangle
=\sum_n c^{(j)}_{n}\!\bigl(n_g-\tfrac12\bigr)\,c^{(i)}_{n}(n_g),
\label{eq:me1}\\
M^-_{ji}&\equiv\langle j,o|e^{-i\hat\varphi/2}|i,e\rangle
=\sum_n c^{(j)}_{n-1}\!\bigl(n_g-\tfrac12\bigr)\,c^{(i)}_{n}(n_g).
\label{eq:me2}
\end{align}
The matrix elements entering Eq.~(\ref{eq:fgr}) then follow as
\begin{equation}
\bigl|\langle j,o|\cos\tfrac{\hat\varphi}{2}|i,e\rangle\bigr|^{2}
=\Bigl(\tfrac{M^+_{ji}+M^-_{ji}}{2}\Bigr)^{2},\qquad
\bigl|\langle j,o|\sin\tfrac{\hat\varphi}{2}|i,e\rangle\bigr|^{2}
=\Bigl(\tfrac{M^+_{ji}-M^-_{ji}}{2}\Bigr)^{2}.
\label{eq:cossin}
\end{equation}
We diagonalize $\hat H(n_g)$ and $\hat H(n_g-\tfrac12)$ in a charge basis with $|n|\le60$ and sum over the lowest ten final states. The rates are already converged for $|n|\le6$.

\subsection{Gap difference and photon-assisted contribution}
\label{sec:gap}

Two quantities in Eqs.~(\ref{eq:albe})--(\ref{eq:Sminus}) remain to be fixed, $\Delta$ and $\delta\Delta$. The gap, $\Delta/h=45.1$\,GHz, follows from the measured critical temperature through the BCS relation (Sec.~\ref{sec:antenna}). The gap difference is determined by matching the measured ratio of the two rates to the model,
\begin{equation}
\frac{\bar\Gamma_p^1}{\bar\Gamma_p^0}=\frac{\sum_j\Gamma_{1j}}{\sum_j\Gamma_{0j}},
\label{eq:ratio}
\end{equation}
where $\bar\Gamma_p^0$ and $\bar\Gamma_p^1$ are the measured rates averaged over $n_g$, and $\Gamma_{ij}$ are given by Eqs.~(\ref{eq:fgr})--(\ref{eq:Sminus}) at $T=T_{\mathrm{MX}}$.  The QP density $x_{\mathrm{qp}}$ enters Eqs.~(\ref{eq:Splus})--(\ref{eq:Sminus}) only as an overall prefactor and therefore cancels in the ratio. For given $T$ and qubit parameters, the right-hand side of Eq.~(\ref{eq:ratio}) is thus a function of $\delta\Delta$ alone. Equation~(\ref{eq:ratio}) assumes that PAPS is negligible in $\bar\Gamma_p^0$, the opposite limit to the one used for $T_{\mathrm{eff}}$ in Table~\ref{table:S1}.

We use the rates measured on Day~15 in Config.~A at $T_{\mathrm{MX}}=17.2$\,mK. At each $n_g$, $\Gamma_p^0$ and $\Gamma_p^1$ are extracted from a linear fit to $\Gamma_p(P_1)$ (Sec.~\ref{sec:ng:meas}), and their averages over $n_g$ give $\bar\Gamma_p^0$ and $\bar\Gamma_p^1$. Table~\ref{tab:gap} lists these averages, the values of $\delta\Delta$ obtained from Eq.~(\ref{eq:ratio}), and the QP density ratio $x^L_{\mathrm{qp}}/x^H_{\mathrm{qp}}$ obtained by inserting $\delta\Delta$ into the equilibrium ratio above. For all three devices $\delta\Delta$ lies between $0.24$ and $0.36$ times $hf_{01}$, within the small-gap-difference regime of Ref.~\cite{nho2026recovery}. QPs therefore cross the junction without thermal activation in $|1\rangle$. The junctions studied here were not intentionally gap-engineered, and the low $\Gamma_p^1$ reported here reflects a low resident QP density rather than a suppressed tunneling channel. Since $\Gamma_p^1$ is not thermally activated, the extracted $x_{\mathrm{qp}}$ is insensitive to errors in $\delta\Delta$. Varying $\delta\Delta/h$ from $0.5$ to $1.2$\,GHz shifts $x_{\mathrm{qp}}$ of Q2 by less than $10\%$. Even Eq.~(1) of the main text, which assumes equal gaps, yields $x_{\mathrm{qp}}\approx5.86\times10^{-11}$, only about $35\%$ below the value in Table~\ref{tab:gap}.

\begin{table}[t]
\caption{Charge-parity switching rates on Day~15 in Config.~A, at $T_{\mathrm{MX}}=17.2$\,mK, the gap difference inferred from their ratio, the resulting ratio of QP densities in the low- and high-gap films, and the QP density in the low-gap film obtained from $\bar\Gamma_p^1$. At each $n_g$, $\Gamma_p^0$ and $\Gamma_p^1$ are the intercept and the $P_1=1$ extrapolation of a linear fit to $\Gamma_p(P_1)$. The bar denotes the average over $n_g$.}
\label{tab:gap}
\begin{ruledtabular}
\begin{tabular}{lcccccc}
 & $\bar\Gamma_p^0$ (Hz) & $\bar\Gamma_p^1$ (Hz) & $\bar\Gamma_p^1/\bar\Gamma_p^0$ & $\delta\Delta/h$ (GHz) & $x^L_{\mathrm{qp}}/x^H_{\mathrm{qp}}$ & \begin{tabular}{@{}c@{}}$x_{\mathrm{qp}}\equiv x^L_{\mathrm{qp}}$\\($\times10^{-11}$)\end{tabular} \\
\colrule
Q1 & $0.130$ & $3.34$  & $25.7$ & $0.97$ & $15.0$ & $14.8$ \\
Q2 & $0.149$ & $2.04$  & $13.7$ & $0.71$ & $7.3$  & $8.93$ \\
Q3 & $0.328$ & $13.03$ & $39.7$ & $1.19$ & $28.3$ & $56.0$ \\
\end{tabular}
\end{ruledtabular}
\end{table}

The PAPS rate in Eq.~(\ref{eq:paps}) does not distinguish between the qubit states. Since breaking a Cooper pair at the junction leaves one QP on each side, it transfers a single electron across the junction, and the PAPS rate for the transition $|i\rangle\to|j\rangle$ involves the same matrix elements as QP tunneling~\cite{houzet2019},
\begin{equation}
\Gamma^{\mathrm{PAPS}}_{ij}\propto\int d\omega\,e_c(\omega)\,\bar n(\omega,T_{\mathrm{rad}})
\left[K_-(E)\bigl|\langle j,o|\cos\tfrac{\hat\varphi}{2}|i,e\rangle\bigr|^{2}
+K_+(E)\bigl|\langle j,o|\sin\tfrac{\hat\varphi}{2}|i,e\rangle\bigr|^{2}\right],
\label{eq:papsij}
\end{equation}
with $e_c$ and $\bar n$ as in Eq.~(\ref{eq:paps}) and $E=\hbar\omega-hf_{ij}$ the energy left to the two QPs. The factors
\begin{equation}
K_\pm(E)=\int_{\Delta}^{E-\Delta-\delta\Delta}d\epsilon\,
\frac{\epsilon(E-\epsilon)\pm\Delta(\Delta+\delta\Delta)}
{\sqrt{\epsilon^{2}-\Delta^{2}}\sqrt{(E-\epsilon)^{2}-(\Delta+\delta\Delta)^{2}}}
\label{eq:Kpm}
\end{equation}
collect the BCS coherence factors and the densities of states of the two electrodes, and vanish below the pair-breaking threshold $E=2\Delta+\delta\Delta$. Here $\epsilon$ and $E-\epsilon$ are the energies of the two QPs, one in each electrode, and the limits of integration enforce $\epsilon\ge\Delta$ and $E-\epsilon\ge\Delta+\delta\Delta$.  The PAPS rate depends on the qubit state. In $|1\rangle$, the transition to $|0\rangle$ adds $hf_{01}$ to the energy left to the two QPs, so that photons with $\hbar\omega\ge2\Delta+\delta\Delta-hf_{01}$ can break a pair, whereas in $|0\rangle$ the photon must supply the full $2\Delta+\delta\Delta$. Since $\bar n$ decreases with frequency, the PAPS rate is larger in $|1\rangle$ than in $|0\rangle$, by a factor set by how steeply $\bar n$ falls near the threshold and hence by $T_{\mathrm{rad}}$. This factor cannot be evaluated for our qubits, however, since the $T_{\mathrm{eff}}$ of Table~\ref{table:S1}, obtained by attributing the entire $\Gamma_p$ to PAPS, is only an upper bound on $T_{\mathrm{rad}}$. In principle, the two contributions can be distinguished through their temperature dependence, since the PAPS rates appear as temperature-independent offsets~\cite{connolly2024coexistence,nho2026recovery}. For junctions with small $\delta\Delta$, such as those in our devices, however, these offsets are difficult to determine~\cite{nho2026recovery}. For the $\delta\Delta$ of Table~\ref{tab:gap}, the reported $x_{\mathrm{qp}}$, obtained by attributing the entire $\bar\Gamma_p^1$ to QP tunneling, is therefore an upper estimate. The QPs created by pair-breaking photons remain in the films as part of the resident population, and their subsequent tunneling is correctly included in $x_{\mathrm{qp}}$~\cite{diamond2022distinguishing}.

\begin{figure}
\centering
\includegraphics[width=\columnwidth]{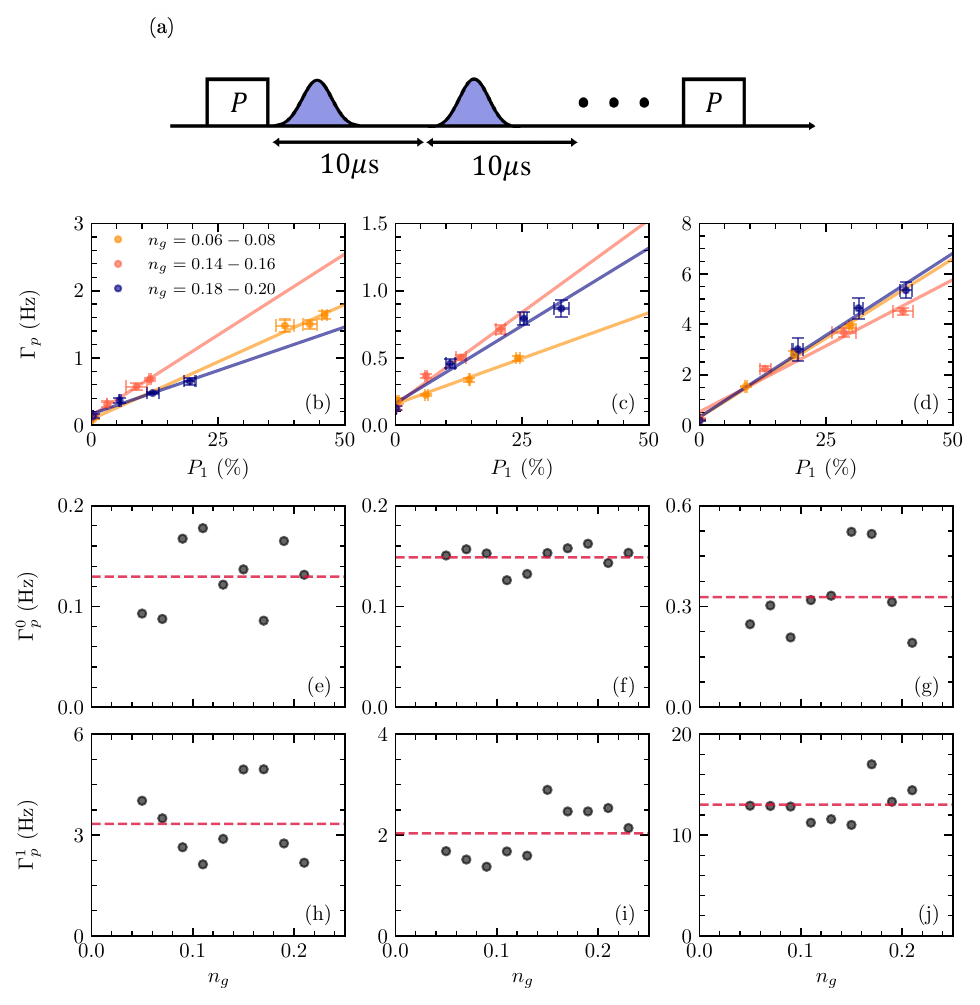}
\caption{Extraction of $\Gamma_{p}^0$ and $\Gamma_{p}^1$. (a) Experimental pulse protocol incorporating variable-amplitude qubit-scrambling pulses. (b)--(d) Measured parity-switching rate $\Gamma_p$ as a function of the excited-state population $P_1$ for Q1--Q3 at various gate charges $n_g$. (e)--(g) Extracted $\Gamma_p^0$ and (h)--(j) extracted $\Gamma_p^1$ as functions of $n_g$ for Q1--Q3. Red dashed lines show the rates predicted by Eqs.~(\ref{eq:fgr})--(\ref{eq:cossin}), with $\delta\Delta$ from Table~\ref{tab:gap} and $x_{\mathrm{qp}}$ chosen to reproduce the average of the measured $\Gamma_p^1$ over $n_g$. The points of Q1 and Q3 at $n_g=0.23$ are omitted, since the linear fits to $\Gamma_p(P_1)$ are poor near $n_g=0.25$, where the parity splitting nearly vanishes. }
\label{fig:S5}
\end{figure}

\subsection{Measured gate-charge dependence}
\label{sec:ng:meas}

Figure \ref{fig:S5} shows measurements of $\Gamma_{p}^0$ and $\Gamma_{p}^1$ for all three qubits, taken on Day~15 in Config.~A. The measurement sequence used to obtain $\Gamma_{p}$ as a function of $P_1$ is illustrated in Fig.~\ref{fig:S5}(a). The excited-state population $P_1$ is controlled using variable-amplitude qubit-scrambling pulses applied every 10\,$\mu$s between successive parity-state measurements \cite{connolly2024coexistence, diamond2022distinguishing,nho2026recovery}. Figures \ref{fig:S5}(b)--(d) show $\Gamma_{p}$ as a function of $P_1$ for Q1--Q3 at various gate charges $n_{g}$, respectively. For all three qubits, $\Gamma_{p}(P_1)$ exhibits a strong dependence on $n_{g}$. This behavior contrasts with previous reports, which found $\Gamma_{p}(P_1)$ to be independent of $n_{g}$ \cite{connolly2024coexistence}.  $\Gamma_{p}^0$ and $\Gamma_{p}^1$ are extracted as the intercept and the $P_1 = 1$ extrapolation of $\Gamma_{p}(P_1)$. The resulting $\Gamma_{p}^0$ and $\Gamma_{p}^1$ as functions of $n_{g}$ for Q1--Q3 are shown in Figs.~\ref{fig:S5}(e)--(g) and \ref{fig:S5}(h)--(j), respectively. 

The red dashed lines in Figs.~\ref{fig:S5}(e)--(j) show the rates predicted by Eqs.~(\ref{eq:fgr})--(\ref{eq:cossin}) with the values of $\delta\Delta$ in Table~\ref{tab:gap}. For all three qubits, the relative variations of these predicted rates over the measured range of $n_g$ are below $2\times10^{-5}$. In contrast, the measured rates vary substantially with $n_g$. Since $x_{\mathrm{qp}}$ enters only as an overall prefactor, it shifts the predicted rates but does not change their relative variation with $n_g$. The observed $n_g$ dependence therefore cannot originate from the matrix elements in Eq.~(\ref{eq:fgr}) that govern QP tunneling. 

A possible origin of this discrepancy is leakage to higher transmon levels, induced by the scrambling pulses or by measurement-induced state transitions (MIST) during readout~\cite{Dumas2024prx,fechant2025offset,WeiEnMIST}. Generalizing the two-level relation of Refs.~\cite{connolly2024coexistence,nho2026recovery} to all transmon levels, the measured $\Gamma_p$ is a population-weighted sum,
\begin{align}
\Gamma_p&=\sum_k P_k\,\Gamma_p^k,
\label{eq:gammap}\\
\Gamma_p^k&=\sum_j\bigl(\Gamma_{kj}+\Gamma^{\mathrm{PAPS}}_{kj}\bigr),
\label{eq:gammak}
\end{align}
where $P_k$ is the population of level $k$ and $\Gamma_p^k$ the parity-switching rate in $|k\rangle$, which contains the QP tunneling and PAPS rates of Eqs.~(\ref{eq:fgr}) and (\ref{eq:papsij}). When only $|0\rangle$ and $|1\rangle$ are occupied, Eq.~(\ref{eq:gammap}) reduces to the linear relation used to extract $\Gamma_p^0$ and $\Gamma_p^1$. For the highly excited levels reached by leakage, $\Gamma_p^k$ exceeds $\Gamma_p^1$ and depends on $n_g$, since the charge dispersion increases exponentially with the level index~\cite{koch2007charge}. Since both the pulse-induced leakage and MIST depend on $n_g$, and MIST also on the initial state~\cite{fechant2025offset}, leakage can produce abrupt changes, such as the step in Fig.~\ref{fig:S5}(i), at different $n_g$ in $\Gamma_p^0$ and $\Gamma_p^1$. The readout also affects the extraction of the rates through $P_1$. Relaxation during readout lowers the fitted $P_1$, and MIST moves part of the population outside the two Gaussian peaks used in the fit, with an effect that varies with $n_g$. Both errors shift the intercept and the $P_1=1$ extrapolation of the linear fit, and an $n_g$-dependent error, such as that from MIST, means that part of the observed variation may reflect the determination of $P_1$ rather than the parity-switching rates themselves.

\section{Parity-switching rate with and without HERD-2 filters}\label{sec:P_CD}

\begin{figure}
\centering
\includegraphics[width=\columnwidth]{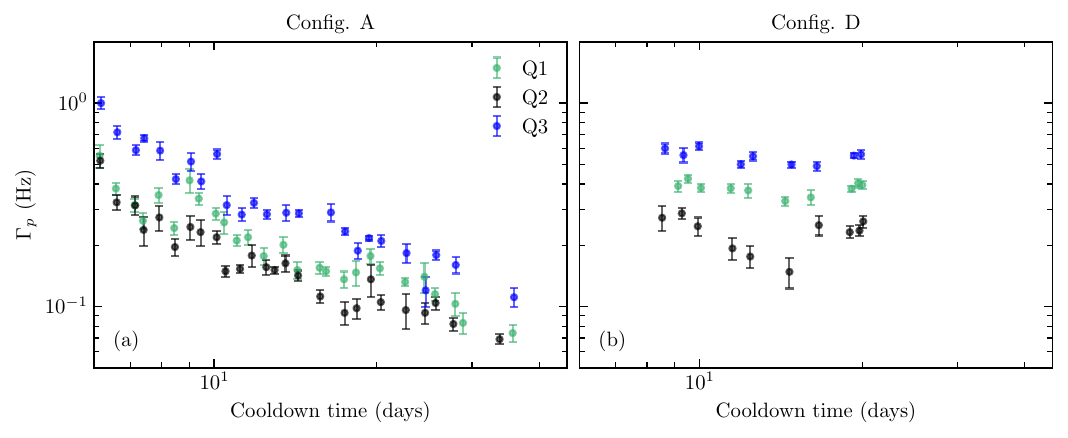}
\caption{Parity-switching rate $\Gamma_p$ of Q1--Q3 as a function of cooldown time in (a) Config.~A, with the HERD-2 filters in place, and (b) Config.~D, with all four HERD-2 filters removed. $\Gamma_p$ continues to decrease in Config.~A but saturates in Config.~D. }
\label{fig:S6}
\end{figure}

In Config.~D, all four HERD-2 filters on the input and output lines were removed, leaving only the Eccosorb filter on the input and no specified IR attenuation on the output, with the enclosure, shielding and the filters on the unused feedthroughs otherwise identical to Config.~A. Figure~\ref{fig:S6} compares $\Gamma_p$ over the two configurations. In contrast to Config.~A, in which $\Gamma_p$ continued to decrease over the cooldown, the switching rates in Config.~D saturated at $0.40$, $0.26$ and $0.56$\,Hz for the three qubits. These saturated rates exceed the lowest rates reached in Config.~A by a factor of about 4--5. Since $\Gamma_p$ in Config.~A was still decreasing, this factor is a lower bound on the ratio of the stationary rates in the two configurations. A rate that no longer decreases indicates a stationary source, consistent with pair-breaking photons propagating along the unfiltered lines.

With the filters in place, the photon flux along the lines is attenuated by a factor $\mathcal{A}$. Since $\Gamma_p^{\mathrm{PAPS}}$ is linear in the incident photon flux, Eq.~(\ref{eq:paps}), the contribution of the filtered lines is at most the rate measured in Config.~D divided by $\mathcal{A}$. Taking $\mathcal{A}$ from the $65$\,dB specified for a single filter over $70$--$120$\,GHz and the largest of these rates, $0.56$\,Hz, and assuming that the attenuation does not decrease above $120$\,GHz, we obtain at most $2\times10^{-7}$\,Hz, more than five orders of magnitude below the lowest rates reached in Config.~A. This is a conservative estimate, since two filters are in series on each path, although their total attenuation need not be their sum once leakage at the connectors and residual radiative paths are taken into account. In-line filtering is therefore required, but once it is in place the coaxial lines no longer limit $\Gamma_p$, and the remaining parity-switching rate originates from the time-dependent source discussed in Sec.~III of the main text.

These results also indicate how the line contribution constrains the number of coaxial lines that can be connected to the package. Assuming that this contribution grows linearly with the number of lines and that each line carries, on average, the same photon flux as the two lines in Config.~D, each line without HERD-2 filters adds $0.10$--$0.22$\,Hz to $\Gamma_p$. This value is obtained by dividing the excess of the saturated rate in Config.~D over the lowest rate reached in Config.~A by the two lines, and is a lower bound since $\Gamma_p$ in Config.~A was still decreasing. From the rates in Config.~A, at most $4$--$10$ lines without HERD-2 filters would thus raise $\Gamma_p$ to $1$\,Hz. With the HERD-2 filters, the contribution of each line falls below $10^{-7}$\,Hz, and $\Gamma_p$ would reach $1$\,Hz only with of order $10^{7}$ filtered lines. As long as every line is filtered and the feedthroughs and connectors are light-tight, the number of lines therefore sets no practical limit on the parity-switching rate.

In-line filters can themselves become a source of radiation if their absorbers are poorly thermalized, and increased switching rates attributed to this effect have been reported~\cite{nho2026recovery}. The filters used here are thermally anchored as described in Sec.~\ref{sec:Experimental setup}. The increase upon their removal shows that they reduce the net pair-breaking photon flux at the device, although a smaller contribution from their own emission cannot be excluded. This increase is small compared with that observed in Config.~B without the MXC shield, where $\Gamma_p$ reaches $16$--$37$\,Hz (Table~II of the main text). The enclosure and shielding described in Sec.~II of the main text therefore provide most of the suppression, and the in-line filters are required to reach the $0.1$\,Hz level.

\section{Effective qubit temperature without HERD-2 filters}\label{sec:Tq_D1}

\begin{figure}
\centering
\includegraphics[width=\columnwidth]{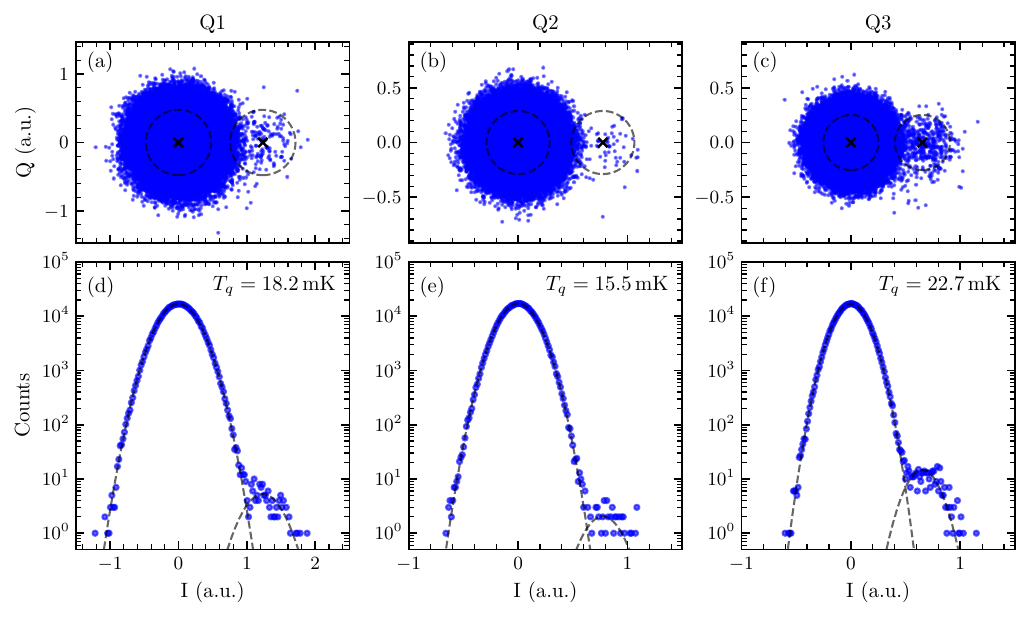}
\caption{Single-shot readout of Q1--Q3 in Config.~D, with 500k shots per qubit. (a)--(c) Shots in the IQ plane, rotated such that the line of centers of the two clusters lies along the I axis. Crosses mark the fitted centers of the ground- and excited-state clusters, and dashed circles have a radius of $2\sigma$. (d)--(f) Histograms of the I quadrature with the double-Gaussian fit (dashed lines). The effective qubit temperature obtained from the fitted excited-state population is indicated in each panel.}
\label{fig:S7}
\end{figure}

We measure the effective qubit temperature $T_q$ in Config.~D from 500k single-shot readouts of each qubit left in thermal equilibrium (Fig.~\ref{fig:S7}). The in-phase (I) and quadrature (Q) components of the readout signal are rotated such that the ground-state cluster lies at the origin and the line of centers of the two clusters lies along the I axis, as shown in Figs.~\ref{fig:S7}(a)--(c). The histogram of I is fitted to the sum of two Gaussians of equal width centered at the two clusters, as shown in Figs.~\ref{fig:S7}(d)--(f), which yields the excited-state population $P_1$. Assuming that the populations of $|0\rangle$ and $|1\rangle$ follow a Boltzmann distribution and neglecting the population of higher levels, the effective qubit temperature follows from
\begin{equation}
T_q=\frac{hf_{01}}{k_B\ln\bigl[(1-P_1)/P_1\bigr]}.
\label{eq:Tq}
\end{equation}
We obtain $P_1=0.030\%$, $0.011\%$ and $0.082\%$, corresponding to $T_q=18.2$, $15.5$ and $22.7$\,mK for Q1--Q3. These values are comparable to those of Configs.~A, B and F, although $T_{\mathrm{MX}}$ in Config.~D is lower (Table~II of the main text). In all configurations except Config.~E, $T_q$ remains between $15$ and $24$\,mK even when $T_{\mathrm{MX}}$ is reduced to about $10$\,mK, indicating that the residual excited-state population is dominated by excitation sources other than the thermal environment at $T_{\mathrm{MX}}$.

\section{Energy relaxation time versus cooldown time}\label{sec:T1_CD}

\begin{figure}
\centering
\includegraphics[width=\columnwidth]{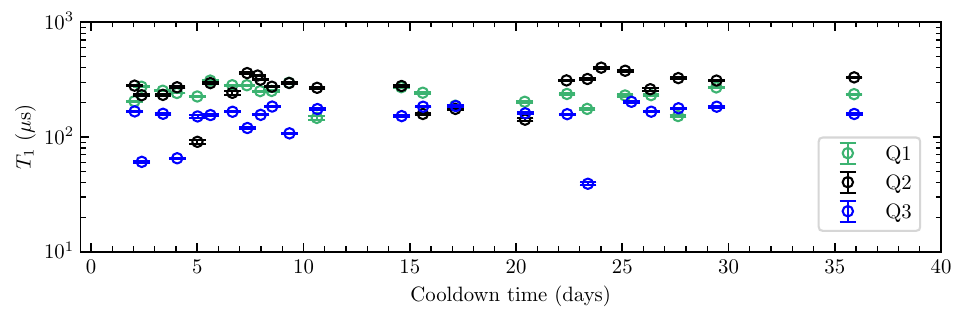}
\caption{Energy relaxation time $T_1$ of Q1--Q3 as a function of cooldown time in Config.~A.}
\label{fig:S8}
\end{figure}

Figure~\ref{fig:S8} shows the energy relaxation time $T_1$ of the three qubits, measured at intervals in Config.~A. Each point is an average over repeated measurements, and the highest of these averages lies between $200$ and $400\,\mu$s for the three qubits. Apart from occasional sudden jumps and statistical fluctuations, $T_1$ shows no significant trend over the cooldown, and the fluctuations of the three qubits are uncorrelated. This behavior is consistent with fluctuating two-level-system defects local to each qubit~\cite{klimov2018fluctuations,schlor2019correlating}. The absence of a trend also indicates that $T_1$ is not limited by QPs. Relaxation accompanied by QP tunneling, $\Gamma_{10}$, accounts for most of $\Gamma_p^1$ and therefore occurs on Day~15 at $2$--$13$\,Hz (Table~\ref{tab:gap}), at least two orders of magnitude below $1/T_1$, and at lower rates later in the cooldown. The decrease of $\Gamma_p$ over the cooldown is thus not expected to affect $T_1$.

\clearpage
\bibliography{SM_citation}